\documentstyle[12pt]{article}

\expandafter\ifx\csname amssym.def\endcsname\relax \else\endinput\fi
\expandafter\edef\csname amssym.def\endcsname{%
       \catcode`\noexpand\@=\the\catcode`\@\space}
\catcode`\@=11
\def\undefine#1{\let#1\undefined}
\def\newsymbol#1#2#3#4#5{\let\next@\relax
 \ifnum#2=\@ne\let\next@\msafam@\else
 \ifnum#2=\tw@\let\next@\msbfam@\fi\fi
 \mathchardef#1="#3\next@#4#5}
\def\mathhexbox@#1#2#3{\relax
 \ifmmode\mathpalette{}{\m@th\mathchar"#1#2#3}%
 \else\leavevmode\hbox{$\m@th\mathchar"#1#2#3$}\fi}
\def\hexnumber@#1{\ifcase#1 0\or 1\or 2\or 3\or 4\or 5\or 6\or 7\or 8\or
 9\or A\or B\or C\or D\or E\or F\fi}

\ifcase\@ptsize
  \font\tenmsa=msam10
  \font\sevenmsa=msam7
  \font\fivemsa=msam5
\or
  \font\tenmsa=msam10  scaled \magstephalf
  \font\sevenmsa=msam7 scaled \magstephalf
  \font\fivemsa=msam5  scaled \magstephalf
\or
  \font\tenmsa=msam10  scaled \magstep1
  \font\sevenmsa=msam7 scaled \magstep1
  \font\fivemsa=msam5  scaled \magstep1
\fi

\newfam\msafam
\textfont\msafam=\tenmsa
\scriptfont\msafam=\sevenmsa
\scriptscriptfont\msafam=\fivemsa
\edef\msafam@{\hexnumber@\msafam}
\mathchardef\dabar@"0\msafam@39
\def\dashrightarrow{\mathrel{\dabar@\dabar@\mathchar"0\msafam@4B}}
\def\dashleftarrow{\mathrel{\mathchar"0\msafam@4C\dabar@\dabar@}}

\def\ulcorner{\delimiter"4\msafam@70\msafam@70 }
\def\urcorner{\delimiter"5\msafam@71\msafam@71 }
\def\llcorner{\delimiter"4\msafam@78\msafam@78 }
\def\lrcorner{\delimiter"5\msafam@79\msafam@79 }
\def\yen{{\mathhexbox@\msafam@55 }}
\def\checkmark{{\mathhexbox@\msafam@58 }}
\def\circledR{{\mathhexbox@\msafam@72 }}
\def\maltese{{\mathhexbox@\msafam@7A }}

\ifcase\@ptsize
  \font\tenmsb=msbm10
  \font\sevenmsb=msbm7
  \font\fivemsb=msbm5
\or
  \font\tenmsb=msbm10  scaled \magstephalf
  \font\sevenmsb=msbm7 scaled \magstephalf
  \font\fivemsb=msbm5  scaled \magstephalf
\or
  \font\tenmsb=msbm10  scaled \magstep1
  \font\sevenmsb=msbm7 scaled \magstep1
  \font\fivemsb=msbm5  scaled \magstep1
\fi

\newfam\msbfam
\textfont\msbfam=\tenmsb
\scriptfont\msbfam=\sevenmsb
\scriptscriptfont\msbfam=\fivemsb
\edef\msbfam@{\hexnumber@\msbfam}
\def\Bbb#1{{\fam\msbfam\relax#1}}
\def\widehat#1{\setbox\z@\hbox{$\m@th#1$}%
 \ifdim\wd\z@>\tw@ em\mathaccent"0\msbfam@5B{#1}%
 \else\mathaccent"0362{#1}\fi}
\def\widetilde#1{\setbox\z@\hbox{$\m@th#1$}%
 \ifdim\wd\z@>\tw@ em\mathaccent"0\msbfam@5D{#1}%
 \else\mathaccent"0365{#1}\fi}

\ifcase\@ptsize
  \font\teneufm=eufm10
  \font\seveneufm=eufm7
  \font\fiveeufm=eufm5
\or
  \font\teneufm=eufm10   scaled \magstephalf
  \font\seveneufm=eufm7  scaled \magstephalf
  \font\fiveeufm=eufm5   scaled \magstephalf
\or
  \font\teneufm=eufm10   scaled \magstep1
  \font\seveneufm=eufm7  scaled \magstep1
  \font\fiveeufm=eufm5   scaled \magstep1
\fi

\newfam\eufmfam
\textfont\eufmfam=\teneufm
\scriptfont\eufmfam=\seveneufm
\scriptscriptfont\eufmfam=\fiveeufm
\def\frak#1{{\fam\eufmfam\relax#1}}

\csname amssym.def\endcsname

\expandafter\ifx\csname pre amssym.tex at\endcsname\relax \else \endinput\fi
\expandafter\chardef\csname pre amssym.tex at\endcsname=\the\catcode`\@
\catcode`\@=11
\newsymbol\boxdot 1200
\newsymbol\boxplus 1201
\newsymbol\boxtimes 1202
\newsymbol\square 1003
\newsymbol\blacksquare 1004
\newsymbol\centerdot 1205
\newsymbol\lozenge 1006
\newsymbol\blacklozenge 1007
\newsymbol\circlearrowright 1308
\newsymbol\circlearrowleft 1309
\undefine\rightleftharpoons
\newsymbol\rightleftharpoons 130A
\newsymbol\leftrightharpoons 130B
\newsymbol\boxminus 120C
\newsymbol\Vdash 130D
\newsymbol\Vvdash 130E
\newsymbol\vDash 130F
\newsymbol\twoheadrightarrow 1310
\newsymbol\twoheadleftarrow 1311
\newsymbol\leftleftarrows 1312
\newsymbol\rightrightarrows 1313
\newsymbol\upuparrows 1314
\newsymbol\downdownarrows 1315
\newsymbol\upharpoonright 1316
 
\newsymbol\downharpoonright 1317
\newsymbol\upharpoonleft 1318
\newsymbol\downharpoonleft 1319
\newsymbol\rightarrowtail 131A
\newsymbol\leftarrowtail 131B
\newsymbol\leftrightarrows 131C
\newsymbol\rightleftarrows 131D
\newsymbol\Lsh 131E
\newsymbol\Rsh 131F
\newsymbol\rightsquigarrow 1320
\newsymbol\leftrightsquigarrow 1321
\newsymbol\looparrowleft 1322
\newsymbol\looparrowright 1323
\newsymbol\circeq 1324
\newsymbol\succsim 1325
\newsymbol\gtrsim 1326
\newsymbol\gtrapprox 1327
\newsymbol\multimap 1328
\newsymbol\therefore 1329
\newsymbol\because 132A
\newsymbol\doteqdot 132B
 
\newsymbol\triangleq 132C
\newsymbol\precsim 132D
\newsymbol\lesssim 132E
\newsymbol\lessapprox 132F
\newsymbol\eqslantless 1330
\newsymbol\eqslantgtr 1331
\newsymbol\curlyeqprec 1332
\newsymbol\curlyeqsucc 1333
\newsymbol\preccurlyeq 1334
\newsymbol\leqq 1335
\newsymbol\leqslant 1336
\newsymbol\lessgtr 1337
\newsymbol\backprime 1038
\newsymbol\risingdotseq 133A
\newsymbol\fallingdotseq 133B
\newsymbol\succcurlyeq 133C
\newsymbol\geqq 133D
\newsymbol\geqslant 133E
\newsymbol\gtrless 133F
\newsymbol\sqsubset 1340
\newsymbol\sqsupset 1341
\newsymbol\vartriangleright 1342
\newsymbol\vartriangleleft 1343
\newsymbol\trianglerighteq 1344
\newsymbol\trianglelefteq 1345
\newsymbol\bigstar 1046
\newsymbol\between 1347
\newsymbol\blacktriangledown 1048
\newsymbol\blacktriangleright 1349
\newsymbol\blacktriangleleft 134A
\newsymbol\vartriangle 134D
\newsymbol\blacktriangle 104E
\newsymbol\triangledown 104F
\newsymbol\eqcirc 1350
\newsymbol\lesseqgtr 1351
\newsymbol\gtreqless 1352
\newsymbol\lesseqqgtr 1353
\newsymbol\gtreqqless 1354
\newsymbol\Rrightarrow 1356
\newsymbol\Lleftarrow 1357
\newsymbol\veebar 1259
\newsymbol\barwedge 125A
\newsymbol\doublebarwedge 125B
\undefine\angle
\newsymbol\angle 105C
\newsymbol\measuredangle 105D
\newsymbol\sphericalangle 105E
\newsymbol\varpropto 135F
\newsymbol\smallsmile 1360
\newsymbol\smallfrown 1361
\newsymbol\Subset 1362
\newsymbol\Supset 1363
\newsymbol\Cup 1264
 
\newsymbol\Cap 1265
 
\newsymbol\curlywedge 1266
\newsymbol\curlyvee 1267
\newsymbol\leftthreetimes 1268
\newsymbol\rightthreetimes 1269
\newsymbol\subseteqq 136A
\newsymbol\supseteqq 136B
\newsymbol\bumpeq 136C
\newsymbol\Bumpeq 136D
\newsymbol\lll 136E
 
\newsymbol\ggg 136F
 
\newsymbol\circledS 1073
\newsymbol\pitchfork 1374
\newsymbol\dotplus 1275
\newsymbol\backsim 1376
\newsymbol\backsimeq 1377
\newsymbol\complement 107B
\newsymbol\intercal 127C
\newsymbol\circledcirc 127D
\newsymbol\circledast 127E
\newsymbol\circleddash 127F
\newsymbol\lvertneqq 2300
\newsymbol\gvertneqq 2301
\newsymbol\nleq 2302
\newsymbol\ngeq 2303
\newsymbol\nless 2304
\newsymbol\ngtr 2305
\newsymbol\nprec 2306
\newsymbol\nsucc 2307
\newsymbol\lneqq 2308
\newsymbol\gneqq 2309
\newsymbol\nleqslant 230A
\newsymbol\ngeqslant 230B
\newsymbol\lneq 230C
\newsymbol\gneq 230D
\newsymbol\npreceq 230E
\newsymbol\nsucceq 230F
\newsymbol\precnsim 2310
\newsymbol\succnsim 2311
\newsymbol\lnsim 2312
\newsymbol\gnsim 2313
\newsymbol\nleqq 2314
\newsymbol\ngeqq 2315
\newsymbol\precneqq 2316
\newsymbol\succneqq 2317
\newsymbol\precnapprox 2318
\newsymbol\succnapprox 2319
\newsymbol\lnapprox 231A
\newsymbol\gnapprox 231B
\newsymbol\nsim 231C
\newsymbol\ncong 231D
\newsymbol\diagup 231E
\newsymbol\diagdown 231F
\newsymbol\varsubsetneq 2320
\newsymbol\varsupsetneq 2321
\newsymbol\nsubseteqq 2322
\newsymbol\nsupseteqq 2323
\newsymbol\subsetneqq 2324
\newsymbol\supsetneqq 2325
\newsymbol\varsubsetneqq 2326
\newsymbol\varsupsetneqq 2327
\newsymbol\subsetneq 2328
\newsymbol\supsetneq 2329
\newsymbol\nsubseteq 232A
\newsymbol\nsupseteq 232B
\newsymbol\nparallel 232C
\newsymbol\nmid 232D
\newsymbol\nshortmid 232E
\newsymbol\nshortparallel 232F
\newsymbol\nvdash 2330
\newsymbol\nVdash 2331
\newsymbol\nvDash 2332
\newsymbol\nVDash 2333
\newsymbol\ntrianglerighteq 2334
\newsymbol\ntrianglelefteq 2335
\newsymbol\ntriangleleft 2336
\newsymbol\ntriangleright 2337
\newsymbol\nleftarrow 2338
\newsymbol\nrightarrow 2339
\newsymbol\nLeftarrow 233A
\newsymbol\nRightarrow 233B
\newsymbol\nLeftrightarrow 233C
\newsymbol\nleftrightarrow 233D
\newsymbol\divideontimes 223E
\newsymbol\varnothing 203F
\newsymbol\nexists 2040
\newsymbol\Finv 2060
\newsymbol\Game 2061
\newsymbol\mho 2066
\newsymbol\eth 2067
\newsymbol\eqsim 2368
\newsymbol\beth 2069
\newsymbol\gimel 206A
\newsymbol\daleth 206B
\newsymbol\lessdot 236C
\newsymbol\gtrdot 236D
\newsymbol\ltimes 226E
\newsymbol\rtimes 226F
\newsymbol\shortmid 2370
\newsymbol\shortparallel 2371
\newsymbol\smallsetminus 2272
\newsymbol\thicksim 2373
\newsymbol\thickapprox 2374
\newsymbol\approxeq 2375
\newsymbol\succapprox 2376
\newsymbol\precapprox 2377
\newsymbol\curvearrowleft 2378
\newsymbol\curvearrowright 2379
\newsymbol\digamma 207A
\newsymbol\varkappa 207B
\newsymbol\Bbbk 207C
\newsymbol\hslash 207D
\undefine\hbar
\newsymbol\hbar 207E
\newsymbol\backepsilon 237F
\catcode`\@=\csname pre amssym.tex at\endcsname

\def\Box{\hbox{\vrule height1ex\kern-0.4pt
\vbox to 1ex{\hrule width1ex\vfil\hrule width1ex}\kern-0.4pt\vrule height1ex}}
\newcommand{\sqr}[2]{{{\vcenter{\vbox{\hrule height.#2pt
\hbox{\vrule width.#2pt height#1pt \kern#1pt
\vrule width.#2pt}
\hrule height.#2pt}}}}}

\newtheorem{prop}{Proposition}
\newtheorem{theorem}{Theorem}
\newcommand{\ovl}{\overline}
\newcommand{\til}{\tilde}
\newtheorem{lemma}{Lemma}

\newcommand{\eo}{\setcounter{equation}{0}}

\newcommand{\be}{\begin{equation}}

\newcommand{\ee}{\end{equation}}
\newcommand{\al}{\alpha}

\newcommand{\dl}{\delta}

\newcommand{\lm}{\lambda}

\newcommand{\rh}{\rho}

\newcommand{\ph}{\phi}

\newcommand{\phv}{\varphi}
\newcommand{\ch}{\chi}
\newcommand{\ps}{\psi}
\newcommand{\Ps}{\Psi}
\newcommand{\om}{\omega}
\newcommand{\Om}{\Omega}

\newcommand{\raw}{\rightarrow}
\newcommand{\A}{\frak A}
\newcommand{\B}{\frak B}
\newcommand{\F}{\frak F}
\newcommand{\C}{{\Bbb C}}

\newcommand{\bib}{\bibitem}
\newcommand{\cin}{C^{\infty}}

\renewcommand{\H}{\mbox{$\cal H$}}

\newcommand{\n}{\parallel}
\newcommand{\h}{{\bf h}}

 \renewcommand{\ll}{\label}

\newcommand{\R}{{\Bbb R}}

\newcommand{\notp}{p \kern-.48em /}
\newcommand{\ci}{\cite}
\newcommand{\bea}{\begin{eqnarray}}
\newcommand{\eea}{\end{eqnarray}}
\newcommand{\ot}{\otimes}
\newcommand{\half}{\mbox{\footnotesize $\frac{1}{2}$}}

\topmargin = - 0.5 cm
\newcommand{\jlt}{\tilde{J_L}}
\newcommand{\enp}{\hfill $\blacksquare$}
\renewcommand{\F}{{\cal F}}
\renewcommand{\L}{{\sf L}}

\newcommand{\wmn}{({\sf m,n})}
\newcommand{\pmn}{\pi^{({\sf m,n})}}
\newcommand{\pim}{\pi^{{\sf m}}}
\newcommand{\pin}{\overline{\pi}^{{\sf  n}}}
\newcommand{\Hmn}{{\cal H}^{({\sf m,n})}}
\newcommand{\Hm}{{\cal H}^{{\sf m}}}
\newcommand{\Hn}{\overline{{\cal H}}^{{\sf n}}}

\newcommand{\plm}{\pi_{{\sf m}}}
\newcommand{\pll}{\pi_{{\sf l}}}
\newcommand{\plla}{\pi_{{\sf l}'}}

\newcommand{\Hlmn}{{\cal H}_{({\sf m,n})}}
\newcommand{\Hlm}{{\cal H}_{{\sf m}}}
\newcommand{\Hll}{{\cal H}_{{\sf l}}}
\newcommand{\Hul}{{\cal H}^{{\sf l}}}

\newcommand{\lu}{{\bf u}_0({\cal H})^*}
\newcommand{\Omn}{{\cal O}_{\sf m,n}}
\newcommand{\pp}{\pi_{\rm pre}}
\newcommand{\pq}{\pi_{\rm qua}}
\newcommand{\Hq}{{\cal H}_{\rm qua}}
\newcommand{\pl}{\hbar}
\newcommand{\cpn}{{\Bbb CP}^n}
\newcommand{\rip}{(\cdot,\cdot)_0}
\newcommand{\pco}{\pi_{\rm co}}
\newcommand{\m}{{\bf m}}
\newcommand{\ghh}{\Gamma_{\rm hol}({\sf H})}

\newcommand{\res}{\upharpoonright}
\newcommand{\plc}{\pi_{\chi}}
\newcommand{\puc}{\pi^{\chi}}
\newcommand{\hlc}{{\cal H}_{\chi}}

\newcommand{\huc}{{\cal H}^{\chi}}
\newcommand{\la}{\langle}
\newcommand{\ra}{\rangle}
\newcommand{\g}{{\bf g}}
\newcommand{\rep}{representation}
\newcommand{\irep}{irreducible representation}

\newcommand{\MW}{Marsden-Weinstein}
\newcommand{\cn}{{\Bbb C}^{n+1}}
\renewcommand{\O}{{\cal O}}
\newcommand{\PH}{{\Bbb P}{\cal H}}
\newcommand{\I}{{\Bbb I}}
\newcommand{\Lk}{\overline{\cal L}_k}
\begin{document}
\setlength{\baselineskip}{1.5\baselineskip}
\thispagestyle{empty}
\title{The infinite unitary group, Howe dual pairs, and the quantization of
constrained systems}
\author{ N.P.~Landsman\thanks{Alexander von Humboldt Fellow and  S.E.R.C.
Advanced Research Fellow}
 \\ \mbox{}\hfill \\ II. Institut f\"{u}r
Theoretische Physik, Universit\"{a}t Hamburg\\ Luruper Chaussee 149, 22761
Hamburg, Germany\\ and \\
 Department of Applied Mathematics and Theoretical Physics,\\ University of
Cambridge, Silver Street, Cambridge CB3 9EW, U.K.\thanks{Present address}}
  \maketitle
\begin{abstract}
The irreducible unitary \rep s of the Banach Lie group $U_0(\H)$ (which is the
norm-closure of the
inductive limit $\cup_k U(k)$) of unitary operators on a separable Hilbert
space $\H$, which were
found by Kirillov and Ol'shanskii, are reconstructed from quantization theory.
Firstly, the coadjoint
orbits of this group are realized as Marsden-Weinstein symplectic quotients in
the setting of dual
pairs.  Secondly, these quotients are quantized on the basis of the  author's
earlier proposal to
quantize a more general symplectic reduction procedure by means of Rieffel
induction (a technique in
the theory of operator algebras).
 As a warmup, the simplest such orbit, the projective Hilbert space, is first
quantized using geometric quantization, and then again with Rieffel induction.

Reduction and induction have to be performed with either $U(M)$ or $U(M,N)$.
The former case
is straightforward, unless the half-form correction to the (geometric)
quantization of the
unconstrained system is applied. The latter case, in which one induces from
holomorphic
discrete series \rep s, is problematic.
For finite-dimensional $\H=\C^k$, the desired result is only obtained if one
ignores half-forms, and
induces from a \rep, `half' of whose highest weight is shifted by $k$ (relative
to the naive orbit
correspondence).
This presumably poses a problem for any theory
of quantizing constrained systems.
 \end{abstract}
 \newpage
\section{Introduction}
\subsection{\MW\ reduction and constrained systems}
The reduced phase space of a constrained mechanical system \ci{GNH} may often
be written as
a so-called Marsden-Weinstein quotient \ci{MW,AM,GS} of the phase space of the
unconstrained system.
Mathematically, this means that certain complicated symplectic manifolds can be
constructed from
perhaps less complicated ones using a canonical reduction procedure \ci{Mum}.

For example, the complex projective space $\cpn$ (equipped with the usual
K\"{a}hler structure
\ci{GH}) is a \MW\ quotient of $\C^{n+1}$ (whose symplectic form $\om$ is
expressed in terms of
the standard inner product, taken linear in the first entry,  by
$\om(\ps,\phv)=-2\, {\rm Im}\,
(\ps,\phv)$) with respect to the group $U(1)$ \ci{AM}. Namely, $U(1)$
(identified with the unit
circle in the complex plane) acts on  $\C^{n+1}$ as follows: $\exp(i\al)\in
U(1)$ maps $\ps\in
\C^{n+1}$ to $\exp(i\al)\ps$; this action is symplectic, and yields an
equivariant moment map \ci{AM}
$J:\cn\raw {\bf u(1)}^*\equiv \R$ given by $J(\ps)=(\ps,\ps)$. Then $\cpn\simeq
J^{-1}(1)/U(1)$.

More generally, given a suitable
symplectic action of a Lie group $H$ on a symplectic space $S$ one may
construct a moment map
\ci{AM,GS} $J:S\raw \h^*$ (where $\h^*$ is the topological dual of the Lie
algebra $\h$ of $H$).
If $J$   intertwines the $H$-action on $S$ with the co-adjoint action on
$\h^*$, the \MW\ reduced
space at $\mu\in\h^*$ is $S^{\mu}=J^{-1}(\mu)/H_{\mu}$, where $H_{\mu}$ is the
stability group of
$\mu$ under the coadjoint action \ci{AM,GS}. (If $\O_{\mu}$ is the co-adjoint
orbit through $\mu$ one
finds that $S^{\mu}\simeq J^{-1}(\O_{\mu})/H$, so that the reduced space only
depends on the orbit
$\O_{\mu}$.) The reduced space (which is a manifold only under further
assumptions) inherits its
symplectic
 structure from $S$, and this may well be the most efficient way of defining
the symplectic
structure of certain spaces (the example above being a case in point).
{\em Here and in what follows, actions and \rep s are assumed continuous.}

\MW\ reduction is a special case of a more general symplectic reduction
procedure \ci{MiW,Xu,NPL93}.
It was recently proposed \ci{NPL93} that this more general procedure should be
quantized by a
technique from operator algebra theory known as Rieffel induction
\ci{Rie74,FD}. This proposal
entails a new approach to the quantization of constrained mechanical systems,
which so far  has been
succesfully tested in the theory of particles moving in external gravitational
and Yang-Mills fields
\ci{NPL93}, abelian gauge field theories \ci{LW}, and in a comparison with
geometric
quantization \ci{Rob}. The purpose of the present paper is to provide further
examples of this
approach in the context of the quantization of infinite-dimensional  K\"{a}hler
manifolds.
\subsection{Quantum mechanics}
One motivation for choosing this class of examples comes from an intriguing
observation of
Tuynman \ci{Tuy,TW} to the effect that the quantization of quantum mechanics is
quantum mechanics
itself. Namely, the space of pure states of a quantum-mechanical system without
superselection rules
is the projective Hilbert space   ${\Bbb
P}\H$, which as a symplectic manifold may be regarded as the phase space of a
classical system
\ci{Mar74,AM}. The geometric quantization of this phase space then reproduces
the original Hilbert
space $\H$. We will review this argument in some detail in section 2, and
complete it by considering
the quantization of the observables. The key point is that it is the class of
observables and
their associated algebraic structure which distinguishes quantum mechanics from
a possible classical
theory defined on ${\Bbb P}\H$.

In ordinary quantum mechanics, any self-adjoint operator $A$ (assumed bounded
for simplicity) on $\H$
corresponds to an observable. Equivalently, one may define a real-valued
function $f_A$ on $\PH$ by
$f_A([\ps])=(A\ps,\ps)$, where the unit vector $\ps\in\H$ is any lift of
$[\ps]\in\PH$.
We shall find that  these $f_A$ are precisely the functions that are
quantizable (in the sense of
geometric quantization) in the holomorphic (or anti-holomorphic)  polarization
of $\PH$.  Other,
more physical characterizations of these  observables
in the context of   quantum mechanics on $\PH$   will be given in subsection
\ref{qm}.

As mentioned above, the K\"{a}hler manifold $\PH$ may be realized as a \MW\
quotient, and as such it
can be quantized by the Rieffel induction technique. Since this technique
provides an alternative
to geometric quantization (given a quantization  of the unconstrained system
$\H$), it is
interesting to see how the special role played by the quantum-mechanical
observables comes about in
the former approach.
\subsection{Howe dual pairs and the representation theory of the unitary group}
{Moreover, we will apply our techniques to quantize a whole class of
K\"{a}hler
manifolds, namely the collection of   quantizable co-adjoint orbits of the
unitary group $U_0(\H)$,
which consists of all unitary operators $U$ on $\H$ for which $U-{\Bbb
I}$ is compact, and which carries the uniform topology; clearly  $U(\C^k)=U(k)$
for $k<\infty$,
and $U_0(\H)$ is the norm-closure of the
inductive limit $U(\infty)\equiv\cup_k U(k)$.
$\PH$ is one such orbit, and its usual K\"{a}hler symplectic structure
coincides with the Lie-Kirillov
symplectic structure of this orbit.) This application was motivated by
Montgomery's observation
\ci{Mon} (also cf.\ \ci{LMS}) that for finite-dimensional $\H$  some these
orbits
(namely  those characterized by  a collection of positive eigenvalues)
 are \MW\ quotients of $\H\ot \C^M$ with
respect to $U(M)$ for suitable $M$ (which depends on the orbit).
We extend this result (which is a special instance of the theory of classical
dual pairs
\ci{KKS,Ste,Wei83}) to the situation where the eigenvalues may be of either
sign, and also to the
case where $\H$ is infinite-dimensional. In the general case one reduces  with
respect to the group $U(M,N)$.

Apart from its obvious relevance to quantum mechanics, our  special interest in
the
infinite-dimensional separable case was triggered by the Kirillov-Ol'shanskii
classification of all
continuous  \rep s of the Banach Lie group $U_0(\H)$
\ci{Kir1,Ols1}. Note that the Fr\'{e}chet Lie group $U(\H)$ (consisting of all
unitary operators on
$\H$), equipped with the strong operator topology, has the same   \rep\ theory
as $U_0(\H)$, because
all
 \rep s of $U_0(\H)$ are also strongly continuous, and can therefore be
extended to
$U(\H)$. Moreover, $U(\H)$ retopologized with the uniform topology has the same
 irreducible \rep s
on separable Hilbert spaces as the same group equipped with the strong topology
(whose irreducible
\rep\ spaces are automatically separable). (The \rep\ theory of $U(\infty)$
equipped with
the inductive limit topology is much more complicated \ci{Ols4,Boy2} and will
not be discussed here.)

A remarkable aspect of this classification (and also the way it was found) is
that all
irreducible unitary  \rep s
of $U_0(\H)$ may be thought of as  the geometric quantization of certain of its
coadjoint orbits.
However, only the geometric quantization of orbits with positive eigenvalues
may actually be found in
the literature \ci{Boy1}; even this special case is already fairly involved
(cf.\ the
Borel-Weil theory (e.g.,  \ci{BE}) for finite-dimensional $\H$).
It
is this quantization that we will be able to redo, and much simplify, by
regarding the orbit as a
constrained system.
 With our formalism
  we merely have to quantize $\H\ot \C^M$, which at first sight is rather
trivially done by Fock space
techniques, and apply Rieffel induction. This last step is easily carried out
on the basis of
  Weyl's classical results on tensor products and the symmetric group
\ci{Wey,How4}.
If, however, one uses the refined version of geometric quantization that
incorporates half-forms
\ci{Woo}
(leading to corrections that are only finite if $\H=\C^k$, $k<\infty$),
 then quantization and reduction fail to commute, and our method breaks down.

The general case (where the orbit is characterized by eigenvalues of arbitrary
sign) is
considerably  more complicated than the special case of fixed sign.
The `answer' is known, in that it is clear from Kirillov's work \ci{Kir1} which
\rep\ of $U_0(\H)$
(on a specific Hilbert space) forms the quantization of a given quantizable
orbit, regarded as the
reduced phase space. Also, the quantization of the unconstrained system
$S=\H\ot\C^{M+N}$
(with a specific symplectic structure depending on $M$ and $N$) is known
explicitly at least for finite-dimensional $\H=\C^k$: it is the $k$-fold tensor
product of the
metaplectic  (alternatively called `oscillator' or `Segal-Shale-Weil') \rep\
\ci{Fol},
restricted from $Sp(2(N+M),\R)$ to its subgroup $U(M,N)$ (see \ci{SW,Ste,BR}).
This tensor product has
been decomposed by Kashiwara and Vergne \ci{KV}, also cf.\ \ci{How2}. (As in
the compact case, one
has the choice whether or not to incorporate half-forms.)

The best point of view concerning this quantization is
provided by the formalism of Howe dual pairs \ci{How1,How2,How3}, which, as
already remarked in \ci{KKS,Ste,Wei83} (and to some extent  anticipated in
\ci{SW}), neatly emerges as
the quantization of the theory of classical (Weinstein) dual pairs \ci{Wei83}
coming from symplectic
group actions on a vector space.
 (A Howe dual pair is defined as a pair of reductive subgroups of a symplectic
group
$Sp(2n,\R)$  which
are each other's centralizer.)
 The trouble is that the decomposition of the Hilbert space
quantizing $S=\C^k\ot \C^{M+N}$  under $U(k)$ and $U(M,N)$ (which form a Howe
dual pair)   does not
reflect the    decomposition of $S$ under these group actions if $k>M+N$ (which
is the case of
relevance to us, as we are eventually interested in $k=\infty$), cf.\
\ci{Ada2}.
 However, a certain modification of our method will lead
to some success.
 \ll{unit}  }
\subsection{Rieffel induction for group actions}
{Let us close this Introduction by briefly reviewing how Rieffel induction
\ci{Rie74,FD} (in the
version of \ci{NPL93}) specializes to the present context. As we have
mentioned, Rieffel induction
quantizes a much more general symplectic reduction procedure than that of
Marsden-Weinstein; the
specialization of this technique to the quantization of \MW\ reduction is only
a slight
generalization of the well-known Mackey induction technique for groups. The
situation is further
simplified if one  deals with reductions by actions of compact
groups (namely $U(M)$).

It is convenient to start from a {\em right} symplectic action of a connected
Lie group $H$ on $S$,
so that the accompanying moment map $J:S\raw (\h^*)$ is an anti-Poisson
homomorphism w.r.t.\ the
Lie-Kirillov Poisson structure on $\h^*$ \ci{AM,GS} (the latter is most easily
defined in terms of
the linear functions on $\h^*$; each $X\in\h$ defines such a linear function by
evaluation, and the
Poisson bracket on $\cin(\h^*)$ is then determined by $\{X,Y\}=[X,Y]$); we
indicate this by writing
$J:S\raw (\h^*)^-$. We assume that the reduced
space $S^{\mu}\equiv J^{-1}(\O_{\mu})/H$ is a manifold.

We adhere to the point of view that
symplectic spaces are best seen as modules for Poisson algebras, and regard the
symplectic reduction
procedure  as a construction in the \rep\ theory of Poisson algebras
\ci{NPLcq}. Thus we suppose that
a Poisson subalgebra $\sf A$ of $\cin(S)$ is given, whose `induced' \rep\
$\pi^{\mu}$ on $S^{\mu}$ we
wish to construct. A sufficient condition on $\sf A$ allowing this
construction is that  each element of $\sf A$ is $H$-invariant; given that $H$
is
connected, this may be reformulated algebraically by requiring that $\sf A$ lie
in the Poisson
commutant of $J^*(\cin(\h^*))$. (A necessary and sufficient condition
is that each element of $\sf A$ is $H$-invariant on
$J^{-1}(\O_{\mu})$.)
 The Poison algebra homomorphism $\pi^{\mu}:{\sf A}\raw
\cin(S^{\mu})$ is then simply defined by the condition that $pr^*
\pi^{\mu}(f)=f$ on
$J^{-1}(\O_{\mu})$ (here $pr:J^{-1}(\O_{\mu})\raw J^{-1}(\O_{\mu})/H$ is the
canonical projection).
For example, if a Lie group $G$ acts symplectically on $S$ in such a way that
its action commutes
with the $H$-action, one could take ${\sf A}=J_L^*(\cin(\g^*))$ (where
$J_L:S\raw \g^*$ is a moment
map, not necessarily equivariant, corresponding to the $G$-action).

Alternatively, one could forget the Poisson algebra $\sf A$ and simply regard
$S^{\mu}$ as a
symplectic $G$-space in the obvious way; one has then constructed an `induced
symplectic
realization', or `classical \rep', of $G$ itself, rather than of its associated
Poisson algebra. The
well-known symplectic induction procedure \ci{KKS,GS} is a special case of this
construction
(it is obtained by taking $H\subset G$ and $S=T^*G$).
Thus induction and reduction are the same; the former terminology is
more
appropriate when starting from $\O_{\mu}$, whereas the latter is
natural when one has $S$ in mind.
 In the
main text we will take $S=\H\ot\C^{M+N}$, $G=U_0(\H)$, and $H=U(M,N)$ with
their natural left- and
right actions on $S$, respectively.

To quantize the reduced space $S^{\mu}$ and the associated induced \rep\ of
$\sf A$ or $G$, we
assume that a quantization of  the unconstrained system as well as of  the
constraints are given.
In the examples  studied in this paper, the required data specified below are
obvious,
and therefore we will refrain from giving an exact definition of
`quantization'; the term will be
used in a somewhat loose way, and everyone's favourite definition will lead to
the objects we use in
our examples.

 Hence we suppose we have firstly found a Hilbert space $\F$,
 which may be thought of as the (geometric) quantization of $S$ (if
$S=\H\ot\C^M$ we take $\F$ to be
the symmetric Fock space $\exp(S)$  over $S$).  Secondly, a   unitary
right-action (i.e., anti-\rep)
$\pi_R$ on $\F$ should be given,  which is the quantization of the symplectic
right-action of $H$ on
$S$ (for $\F=\exp(S)$ this will be the second quantization of the action on
$S$).
Thirdly, we require a unitary \rep\ $\plc(H)$ on a Hilbert space $\hlc$, which
`quantizes' the
coadjoint action of $H$ on the coadjoint orbit $\O_{\mu}$ This is only possible
if the orbit is
`quantizable'; for $H=U(M)$ there is a bijective correspondence between such
orbits and unitary
 \rep s, and for $U(M,N)$ one obtains at least all unitary highest weight
modules by `quantizing' such
  orbits \ci{Ada1,Vog}. (In the latter case the concept of quantization has to
be stretched somewhat
to incorporate the derived functor technique to construct \rep s.)

First assuming that $H$ is compact,  we construct the induced space $\huc$ from
these data
as the subspace of $\F\ot\hlc$ on which $\pi_R^{-1}\ot\plc$ acts trivially
(here $\pi_R^{-1}$ is the
\rep\ of $H$ defined by $\pi_R^{-1}(h)=\pi_R(h^{-1})$). If $H$ is only locally
compact (and assumed
unimodular for simplicity) with Haar measure $dh$, one has to find a dense
subspace $L\subset \F$
such that the integral $\int_H dh\,( (\pi_R^{-1}\ot\plc)(h)\Psi,\Phi)\equiv
(\Psi,\Phi)_0$ is finite
for all $\Psi,\Phi\in L\ot\hlc$. This defines a sesquilinear form $\rip$ on
$L\ot\hlc$ which can be
shown to be positive semi-definite under suitable conditions \ci{NPL93}. The
induced space $\huc$ is
then defined as the completion of the quotient of $L\ot\hlc$ by the null space
of $\rip$; its inner
product is, of course, given by the quotient of $(\cdot ,\cdot )_0$.
 For $H$ compact the integral exists for
all $\Psi,\Phi\in\F$ and $(\Psi,\Phi)_0=(P_0\Psi,P_0\Phi)$, where $P_0$ is the
projector onto the
subspace of $\F\ot\hlc$ carrying the trivial \rep\ of $H$, so that we recover
the first description
of $\huc$. (Even the case where $H$ is not locally compact can sometimes be
handled by a limiting
procedure, cf.\ \ci{LW}.)

We now assume that a group $G$ or
a $\mbox{}^*$-algebra $\A$ acts on $\F$ through a unitary \rep\ or a
$\mbox{}^*$-\rep\ (which we both
denote by $\pi_L$), respectively; it is required that these actions commute
with $\pi_R(H)$.
The  self-adjoint part of the $\mbox{}^*$-algebra $\A$ is thought of as the
(deformation)
quantization of the Poisson algebra $\sf A$, and the actions of $\A$ or $G$ on
$\F$ should be the
quantum counterparts of the actions of $\sf A$ or $G$ on $S$.
(In our example, the action of $G=U_0(\H)$ on $\F=\exp(\H\ot\C^k)$ is the
second quantization of the
left-action of $U_0(\H)$ on $S=\H\ot\C^k$).

The induced \rep s $\puc(\A)$ or $\puc(G)$ on $\huc$ are now defined as
follows.
For $H$ compact, $\puc$ is simply the restriction of $\pi_L\ot{\Bbb I}$ to
$\huc\subset \F\ot\hlc$;
this is well defined because $\pi_L\ot{\Bbb I}$ commutes with
$\pi_R^{-1}\ot\plc$.
In the general case,
  one has to assume that $\pi_L$ leaves $L$ stable; then $\puc$ is essentially
defined as the quotient
of the action of $\pi_L\ot{\Bbb I}$ (on $L\ot\hlc$) to $\huc$ as defined above
(cf.\ \ci{NPL93} for
technical details pertinent to the general case).
The Mackey induction procedure for group \rep s is recovered by assuming that
$H\subset G$, and
taking $\F=L^2(G)$, cf.\ \ci{Rie74,FD} for details in the original setting of
Rieffel induction, and
\ci{NPL93} for the above setting.

As a simple example, take $\F=L^2(G)$ for a locally compact but non-compact
unimodular  group $G$,
and $H=G$, which act on $\F$ in the left- and right-regular \rep s,
respectively. We induce from
the trivial \rep\ $\pi_{\ch}=\pi_{\rm id}$. We may choose $L=C_c(G)$, and
define $V:L\raw \C$ by
$V\ps=\int_G dx\, \ps(x)$. This integral exists, and
$(V\ps,V\phv)=(\ps,\phv)_0$ (where the
left-hand side is the inner product in $\C$).
Hence we can identify the
null space of $(\cdot ,\cdot )_0$ with the kernel of $V$, and the induced space
$\H^{\rm id}$   with
the
  image of $V$, that is, with $\C$. The induced \rep\ $\pi^{\rm id}(G)$ comes
out to be the trivial
one.  This example illustrates the interesting point that Rieffel induction
does not necessarily
produce \rep s that are weakly contained in $\F\ot\H_{\ch}$. For $G=\R^n$ it so
happens that the
trivial \rep\ is weakly contained in the regular one, but for $G$ semi-simple
(and non-compact) it
is not. Yet Rieffel induction manages to extract it in either case.\ll{ri} }
\section{The geometric quantization of quantum mechanics}
In this section we review (and somewhat elaborate on) Tuynman's argument that
the geometric
quantization of the symplectic formulation of quantum mechanics reproduces the
usual Hilbert space
formalism \ci{Tuy,TW}, and complete the thesis by incorporating the
quantization of the observables.
We will keep the discussion technically simple by assuming that $\H$ is
finite-dimensional
(the infinite-dimensional case will be dealt with later,  using the appropriate
Riefel induction
technology).

The author is indebted to G.M. Tuynman for comments on the first draft
of this paper.

 \subsection{Prequantization}
We assume that the reader is somewhat familiar with the ideas of geometric
quantization
\ci{Woo,AM},		 so we will mainly establish our notation in this subsection.
Interestingly, the
argument runs slightly differently depending on which sign conventions one uses
for Hamiltonian
vector fields.
 We start by using the conventions mostly used by
mathematicians (which, indeed, are the ones employed in \ci{Tuy,TW}). Here  the
 Hamiltonian vector
field $\xi_f$ of $f\in\cin(S)$ is defined by $i_{\xi_f}\om=-df$, where $\om$ is
the symplectic form
on $S$.  Similarly, the generator $f_{\xi}$ of a vector field $\xi$, whose flow
leaves $\om$
invariant,
  is defined
by $i_{\xi}\om=-df_{\xi}$. The Poisson bracket is
$\{f,g\}=\om(\xi_f,\xi_g)=\xi_f g$.
This implies that
$[\xi_f,\xi_g]=\xi_{\{f,g\}}$ and  $\{f_{\xi_1},f_{\xi_2}\}=f_{[\xi_1,\xi_2]}$
(plus a possible
central extension).

In geometric quantization one attempts to find a line bundle $\L$ over $S$ with
connection $A$ and
curvature $F_A$, satisfying
\be
F_A=-\frac{i}{\pl}pr^*\om,\ll{chern}
\ee
where $pr:\L\raw S$ is the canonical projection.
For $\pl=1/2\pi$ this is the condition $c_1(\L)=[\om]$ stating that the Chern
class of the line
bundle equals the cohomology class of the symplectic form, cf.\ \ci{GH}.
For any $f\in\cin(S)$, the prequantization $\pp(f)$ is an (unbounded) operator
defined on the linear
space of smooth sections of $\L$ with compact support; this space has a natural
inner product
derived from the Liouville measure on $S$  (if ${\rm dim}\, S=n$ this measure
corresponds to the
volume form $\om^n$), and the completion may be identified with $L^2(S)$. The
prequantization is
defined by
\be
\pp(f)=\frac{\pl}{i}\nabla_{\xi_f}+f, \ll{pre}
\ee
where $\nabla$ is the covariant derivative defined by the connection $A$, and
$f$ is a
multiplication operator.  The crucial property satisfied by   prequantization
is
\be
[\pp(f),\pp(g)]=-i\pl\pp(\{f,g\}). \ll{dir}
\ee
If a Lie group $G$ acts on $S$, we may define a vector field $\xi_X$ for each
$X\in\g$ by
$(\xi_X f)(s)=d/dt f(\exp(-tX)s)_{t=0}$. Writing $f_X$ for $f_{\xi_X}$, one
finds that $X\raw
f_X$ and $X\raw (i/\pl)\pp(f_X)$ are Lie algebra homomorphisms up to a possible
central extension.
\subsection{Prequantization of $\cpn$}
We will now prequantize the projective space of $\H=\C^{n+1}$. We
choose $S=\PH=\cpn$.
We define its symplectic structure through \MW\ reduction, cf.\ the
Introduction (for a direct
definition cf.\ \ci{GH}). We start from $\H$, equipped  with symplectic form
$\til{\om}(\ps,\phv)=-2\pl\, {\rm Im}\, (\ps,\phv )$. The space $ {\Bbb
S}\H=\{\ps\in\H|(\ps,\ps)=1\}$ is
co-isotropically embedded in $\H$; the quotient by its null foliation is $\PH$,
which consists of
equivalence classes $[\ps]$ in $ {\Bbb S}\H$, where $\ps_1\sim\ps_2$ iff
$\ps_1=z\ps_2$ for some $z\in\C$
with $|z|=1$. The symplectic form $\om$ on $\PH$
is then the reduction of $\til{\om}$, cf.\ \ci{AM,GS}.

Let $pr$ be the canonical projection from $ {\Bbb S}\H$ to $\PH$. This
projection makes  $ {\Bbb S}\H$ a principal
fibre bundle over $\PH$ with structure group $U(1)$.  We denote the generator
of $U(1)$ by $T$
($T=i$ in the defining \rep\ of $U(1)$ on $\C$). The vertical vector $v_T(\ps)$
at $\ps\in  {\Bbb S}\H$ is
$i\ps$, where we have identified $T_{\ps} {\Bbb S}\H\subset T_{\ps}\H\simeq \H$
with a subspace of
$\H$ according to
\be
T_{\ps} {\Bbb S}\H=\{\phv\in\H|{\rm Re}\, (\ps,\phv)=0\}. \ll{iden}
\ee
This bundle carries a connection $A$ defined by
\be
\la A,\phv\ra(\ps)={\rm Im}\, (\phv,\ps)\ot T. \ll{con}
\ee
Clearly, $\la A,v_T\ra (\ps)=T$ as required. It is then clear that the
prequantization line bundle
$\sf L$ is the hyperplane bundle $\sf H$ over $\cpn$ \ci{GH}: this is the line
bundle associated to
the principal bundle
$({\Bbb S}\cn,\cpn,pr)$ by the representation $z\raw \ovl{z}$, or $T\raw -i$,
of $U(1)$.
(For one may extend the tangent vectors $\phv_1,\phv_2\in T_{\ps} {\Bbb S}\H$
to vector fields in
a neighbourhoud of $\ps$ satisfying $[\phv_1,\phv_2]=0$, and then the formula
$\la dA|\phv_1,\phv_2\ra =\phv_1\la A,\phv_2\ra -\phv_2\la A,\phv_1\ra -\la
A,[\phv_1,\phv_2]\ra$ and
the replacement $T\raw -i$ in the definition of $A$ shows that (\ref{chern}) is
satisfied.)
\subsection{Quantization of $\cpn$}
To pass from prequantization to quantization we use the anti-holomorphic
polarization on $\cpn$;
in local co-ordinates this is the distribution  $F$
on $\cpn$  which is  spanned
by $\{\partial/\partial \ovl{z_i}\}_i$.
Since the connection is analytic \ci{GH}, this polarization determines the
polarized sections of $\sf H$ as the holomorphic ones.
The space $\ghh$ of holomorphic sections of $\sf H$ is well known (e.g.,
\ci{GH}): realizing the
sections of $\sf H$ as equivariant functions
\be
\ghh\ni\Psi:{\Bbb S}\cn\raw \C ; \:\:\:\: \Psi(\ps z)=z\Psi(\ps)\ll{equ}
\ee
for all $z\in
U(1)$, the holomorphic ones are in one-to-one correspondence with vectors
$\phv\in \cn$, and given by
$\Psi_{\phv}(\ps)=(\ps,\phv)$. Hence we obtain a linear map $V:
\ovl{\H}\raw\ghh$ from the conjugate
space of $\H$ to the Hilbert space $\Hq=\ghh$ of the  geometric quantization of
$S=\PH$, given by
$(V\phv)(\ps)=(\ps,\phv)$. The inner product on $\ghh$ is given by the
Hermitian structure
of the line budle $\sf H$ and integration over $\PH$ w.r.t.\  the Liouville
measure obtained from the
symplectic form - there is no need for half-densities or so in this case.
Normalizing the   Liouville measure suitably, it follows that the map $V$ is
unitary (note that
the inner product on $\ovl{\H}$ is the complex conjugate of that on $\H$).

The final step in the geometric quantization of $\PH$ (omitted in \ci{Tuy,TW})
is the quantization of
(a subset of) the observables, i.e., the smooth functions on $S=\PH=\cpn$. Only
those functions
$f\in\cin(S)$ are quantizable which  satsify the condition that
$\pp(f)\Psi\in\ghh$ for all
$\Psi\in\ghh$. This is equivalent to the
requirement that $[\xi_f,\xi]\in F$ for all $\xi\in F$. Hence $\xi_f$ generates
a holomorphic
diffeomorphism of $\cpn$ (the vector field is automatically complete because
$\cpn$ is compact).

In a move analogous to the proof of Wigner's
theorem in \ci{TW}, we now use Chow's theorem \ci{GH}, which implies that any
holomorphic
  diffeomorphism of $\cpn$ is induced by an invertible linear transformation of
$\C^n$. If we
realize $\cpn$ as $\C^{n+1}/\C^*$, and denote the corresponding  projection
from $\C^{n+1}$ to $\cpn$
by $pr$, this corollary of Chow's theorem means that $\xi_f(pr(\ps))=-pr_*\,
X\ps$, where $X\in {\bf
gl}_n(\C)$ and $X\ps\in T_{\ps}\C^{n+1}\simeq \cn$. But we know in addition
that $\xi_f$ is the
Hamiltonian vector field of a (real-valued) function in $\cin(\cpn)$; in
particular, $\xi_f$ must
leave the symplectic form invariant. Hence $X^*=-X$, and the flow of $\xi_f$ is
induced by unitary
transformations $\exp(tX)$ of $\cn$. Therefore, $X\ps$ is tangent to $S\cn$,
cf.\ (\ref{iden}), and
we may return to our characterization of $\cpn$ as $S\cn/U(1)$.
 A simple exercise shows that the function producing this $\xi_f$ as its
Hamiltonian vector field is
given by
\be
f_X([\ps])=i\pl(X\ps,\ps), \ll{pqf}
\ee
where $\ps\in\cn$ is an arbitrary preimage of $[\ps]\in\cpn$. Conversely, the
group $G=U(n+1)$ has a
symplectic action on $\cpn$ obtained by projecting its defining action on
$\cn$. For each $X\in {\bf
u}_{n+1}$ the function $f_X$ is then defined as explained after (\ref{dir}).

Before clarifying the significance of the result (\ref{pqf}), we will describe
the quantization
$\pq(f_X)$; this is just the restriction of $\pp(f_X)$ to $\Hq=\ghh$.  With
$pr: {\Bbb S}\H\raw\PH$ we
exploit the fact that $\xi_X(pr(\ps))=-pr_* X\ps$, where $X\ps\in T_{\ps} {\Bbb
S}\H$, cf.\ (\ref{iden}).
With $\Psi\in\ghh$ realized as in (\ref{equ}), the covariant derivative acts
according to
$\nabla_{\xi}\Psi(\ps)=(\til{\xi}-v_{\la A,\til{\xi}\ra})\Ps(\ps)$, where
$\til{\xi}\in T_{\ps} {\Bbb S}\H$
is an arbitrary lift of $\xi\in T_{pr(\ps)}\PH$. With $\xi(\ps)=\xi_X(\ps)$ we
of course choose the
lift $\til{\xi_X}(\ps)=-X\ps$.  Using (\ref{con}) and (\ref{pqf}) one finds
that with this choice
 $\pl \la A, \til{\xi_X}\ra=T\ot f_X$. With (\ref{equ}) and the fact that
$T=-i$ on the
hyperplane bundle $\sf H$, we find that the multiplication operator $f$ in
(\ref{pre}) cancels the
term in $\nabla_{\xi_X}$ that comes from the connection $A$. At the end of the
day we therefore
obtain
\be
(\pq(f_X)\Psi)(\ps)= i\pl\frac{d}{dt}\Psi(e^{tX}\ps)_{t=0}. \ll{day}
\ee

As explained after (\ref{dir}), we can extract a \rep\ of the Lie algebra of
$U(\H)=U(n+1)$, which
in this case exponentiates to a \rep\ $\pq$ of $U(\H)$.
Realized on $\ovl{\H}=V^{-1}\Hq$, we find from (\ref{day}) that
$\pq(U)=\ovl{U}$.

We recall the steps leading to this result: the defining \rep\ of $U(\H)$ on
$\H$ induces a
symplectic action on $\PH$, which is generated by the functions $f_X$. These
can be quantized, which
leads to a \rep\ of ${\bf u}(\H)$, which in turn is exponentiated to
$\pq(U(\H))$. That the latter is
the conjugate of the action on $\H$ we started from was to be expected from the
identification of
$\Hq$ with $\ovl{\H}$. As we shall see in subsection \ref{sign}, this curious
conjugation is merely a
consequence of the sign conventions we have chosen (following \ci{Tuy,TW}).
 \subsection{More on the observables of quantum mechanics}
{The description of the observables  of quantum mechanics as those (smooth)
functions
(\ref{pqf}) on $\PH$ that can be quantized in the anti-holomorphic polarization
may not be their most
compelling characterization. A physically more meaningful property of the
function $f_X$
(where $X\in {\bf u}_{n+1}$)
 is that it  can be extended to an affine
function on the  state space $K$ of the $C^*$-algebra $M_{n+1}(\C)$   of linear
operators on $\H=\cn$.
This state space consists of all normalized  positive linear functionals on
$M_{n+1}(\C)$, hence each
element $\om$ of $K$ satisfies $\om({\Bbb I})=1$ and $\om(A^*A)\geq 0$.
$K$ is a  compact
convex set whose extreme boundary of pure states is the `phase space'  $\cpn$.
The embedding of $\cpn=\PH$ into $K$ is obtained by realizing that  a unit
vector $\Omega\in\H$
defines a state $\om$ by $\om(A)=(A\Omega,\Omega)$.
Each mixed state $\om$ in $K$ admits a (highly nonunique)  extremal
decomposition $\om=\sum_ip_i\om_i$
(with $\sum_i p_i=1$) as a convex sum of pure states $\om_i\in\PH$.

A visually accessible example is provided by $\H=\C^2$, so that $\PH={\Bbb
CP}^1=S^2$. The state
space of $M_2(\C)$ (the algebra of $2\times  2$ matrices) is the unit ball
$B^3$ in $\R^3$; its
extremal boundary, the two-sphere $S^2$ with unit radius, is the pure state
space.
Points in the interior may be writen as convex sums of boundary points in many
ways.

A skew-adjoint operator $X$ defines a
continuous real-valued function $f_X$ on $K$ by $f_X(\om)=i\om(X)$; when
restricted to the pure state
space this function clearly coincides with (\ref{pqf}). Conversely, $f_X\in
C(K)$ is the unique
affine extension of $f_X\in\cin(\PH)$ ( a function $f$ on $K$ is called affine
if $f(\lm\om_1+(1-\lm)\om_2)=\lm f(\om_1)+(1-\lm)f(\om_2)$ for all
$\om_1,\om_2\in K$ and $0<\lm<1$).
An
affine function on $K$ is uniquely determined by its values on $\PH$.
However, a generic function on $\PH$ cannot be extended to an affine function
on $K$,
because different extremal decompositions of a point in $K$ would produce
different values of the
(extended) function at that point. The (relatively few) functions on $\PH$
which are insensitive to
this nonuniqueness are precisely the `linear' observables $f_X$  of quantum
mechanics. On ${\Bbb
PC}^2={\Bbb CP}^1$ there are only four such (linearly independent) observables!
(See \ci{AS76} for the general theory of affine function spaces  on compact
convex sets.)

An alternative  characterization of these observables $f_X$ comes from the
transformations
$\phv_t^X=\exp(t\xi_X)$ of $\PH$ they generate via their Hamiltonian vector
fields $\xi_X$. We have
already seen that $\phv_t^X$ leaves the symplectic as well as the complex (and
thereby the
K\"{a}hler) structure of $\PH$ invariant, cf.\ \ci{Italians}. This implies that
the transition
probability (which on $\H$ is given by $|(\ps,\phv)|^2$, and quotients to
$\PH$) is invariant under
the flow $\phv_t^X$ of $f_X$. Conversely, Wigner's theorem implies that any
transformation of $\PH$
with this property is  generated by a function of the type $f_X$ (possibly
composed with the
anti-symplectic transformation on $\PH$ which is induced from the map
$\ps\raw\ovl{\ps}$ on $\H$),
cf.\ \ci{TW}. A theorem of Shultz \ci{Shu} then allows us to characterize the
observables as
those continuous functions on $\PH$ whose flow is the restriction to the pure
state space $\PH$ of
an affine homeomorphism of the total state space $K$. Finally, the equivalence
of all descriptions
listed is then confirmed by Kadison's theorem \ci{Kad51}
  that any affine homeomorphism $\phv_t$ of the
state space $K$ of a $C^*$-algebra $\A$ is induced by a Jordan morphism of
$\A$; in the present case
$\A=M_{n+1}(\C)$ this implies that $\phv_t$ must be induced by a unitary- or an
anti-unitary operator
on $\H=\cn$.

Note, that the $f_X$ form a subset of the Poisson algebra $\cin(\PH)$, but not
a Poisson subalgebra:
the relevant commutative multiplication is not the pointwise one used in
classical mechanics,
but the Jordan product $f_X\circ f_Y=\half i\hbar f_{XY+YX}$.
This product may be motivated by non-commutative spectral theory on convex sets
\ci{AS76}, or
by considerations involving the K\"{a}hler geometry of the pure state space
\ci{Italians}.

 With the exception of the compactness of $K$, all considerations in this
subsection are equally
well valid for $n=\infty$, if $M_{\infty}(\C)$ is taken to be the $C^*$-algebra
of compact operators.
We see that from a physical point of view it is the affine structure of the
total state space,
rather than the complex structure of the pure state space, which is essential.
 \ll{qm} }
\subsection{New sign conventions} {We will actually recover $\H$ (rather than
$\ovl{\H}$) from the
geometric quantization of $\PH$ if we follow the conventions of \ci{AM}, and
define
 the  Hamiltonian vector field
$\xi_f$ of $f\in\cin(S)$   by $i_{\xi_f}\om=df$.
   The Poisson bracket is now $\{f,g\}=\om(\xi_f,\xi_g)=-\xi_f g$.
This implies that
$[\xi_f,\xi_g]=-\xi_{\{f,g\}}$.
If a Lie group $G$ acts on $S$, we redefine the vector field $\xi_X$ for each
$X\in\g$ by
$(\xi_X f)(s)=d/dt f(\exp(tX)s)_{t=0}$. Then $[\xi_X,\xi_Y]=-\xi_{[X,Y]}$, but,
as with the
old conventions,
$\{f_{X},f_{Y}\}=f_{[X,Y]}$ (plus a possible central extension). For geometric
quantization these
conventions imply that the connection on the prequantization bundle  now has to
satisfy
$F_A=(i/\pl)pr^*\om$, rather than (\ref{chern}). The  prequantization itself is
still given by
(\ref{pre}).
Instead of (\ref{dir}), one now has $[\pp(f),\pp(g)]=i\pl\pp(\{f,g\})$. Hence
we obtain a (projective) \rep\ of $\bf g$ by $X\raw (-i/\pl)\pp(f_X)$.

Since we have not changed the symplectic form $\om$ on $\PH$, the
prequantization bundle is now
obviously the tautological line bundle $\sf T$ over $\PH$ \ci{GH}, which is
associated to the
principal bundle $ {\Bbb S}\H$ over $\PH$ via the defining \rep\ of $U(1)$ on
$\C$.
The space of holomorphic sections of $\sf T$ being empty \ci{GH}, we now choose
the holomorphic
polarization on $\PH$ to go from prequantization to quantization. The
antiholomorphic sections of
$\sf T$ are all of the form $\Psi(\ps)=(\phv,\ps)$ for some $\phv\in\H$, so
that we find a unitary
map $V:\H\raw \Hq$ given by $(V\phv)(\ps)=(\phv,\ps)$.
 The quantization of $f_X$ is given by {\em minus} (\ref{day}). The \rep\ of
$U(\H)$ defined through
this quantization (cf.\ the text following (\ref{day})) is now simply the
defining \rep\ on $\H$.
Hence geometric quantization has indeed recovered $\H$ from $\PH$! \ll{sign}}
\section{Representations of the unitary group from Rieffel induction}\eo
 Our main purpose
in this chapter is to
obtain all irreducible unitary \rep s of the unitary groups $U(\H)$ and
$U_0(\H)$ (cf.\ subsection
\ref{unit}) from an induction construction.
In subsections 3.1 to 3.3 we take $\H$ to be an infinite-dimensional separable
Hilbert space, unless
explicitly stated otherwise. All results (sometimes with self-explanatory
modifications) are equally
well valid in the finite-dimensional case, which is considerably easier to
handle; we leave this to
the reader. We start with the simplest case, the defining \rep, which forms the
bridge between the
preceding part of the paper and what follows.
\subsection{The quantization of $\PH$ revisited}
As explained in the Introduction, we can realize $\PH$ as a \MW\ quotient.
The group $U(1)$ acts on $\H$ (in principle from the right, though this is
irrelevant here)  by
$\ps\raw \ps z$, $\ps\in\H,\, |z|=1$. The most general equivariant moment map
$J_R:\H\raw {\bf
u(1)}^*=\R$ \ci{AM,GS} corresponding to this action is given by
\be
J_R(\ps)=(\ps,\ps)+c, \ll{momm}
\ee
where $c$ is a constant (as explained in subsection \ref{ri},
this is `officially' an anti-Poisson homomorphism, but again this is irrelevant
here).
The reduced space $\H^1=J^{-1}(1+c)/U(1)$ then coincides with with $\PH$ as a
symplectic space
 (that is, including the normalization of the symplectic form). We put $c=0$ in
what follows.

The quantization of this type of reduced space using Rieffel induction  was
outlined in subsection
\ref{ri}. We first need a quantization of the `unconstrained' system $\H$,
which we take to be the
symmetric (bosonic) Fock space $\F=\exp(\H)$ (this is the direct sum of all
symmetrized tensor
products $\H^{\ot n}$ ($n=0,1,\ldots$) of $\H$ with itself). This quantization
is so well-established that
we will not motivate it here; cf.\ \ci{Fol,RS1} for mathematical aspects, and
\ci{Woo} for a derivation in
geometric quantization.

 The (anti) \rep\ $\pi_R$ of $U(1)$ on $\F$ is obtained by `quantization' of
the right
action on $\H$. No physicist would hesitate in choosing  $\pi_R$ as the second
quantization of this
right action. Labelling this choice $\pi_{R,{\rm sq}}$, this yields
 $\pi_{R,{\rm sq}}(z) \res \H^{\ot n}=z^n{\Bbb I}$. Similarly,
the defining \rep\ $\pi_{1}$ of $G=U(\H)$
(the group of all
unitary operators on $\H$)
 on $\H_1=\H$ yields a symplectic action  on $\H$. This is `second' quantized
by the
\rep\ $\pi_{L,{\rm sq}}$ on
$\F$, whose restriction $\pi_n$ to each subspace $\H^{\ot n}\subset\F$ is the
symmetrized $n$-fold
tensor product of $\pi_{1}$ with itself.  The \rep s $\pi_{R,{\rm sq}}(U(1))$
and  $\pi_{L,{\rm
sq}}(U(\H))$ obviously commute with each other. Hence $\F$ has a central
decomposition under
 $\pi_{L,{\rm sq}}(U(\H))\ot \pi^{-1}_{R,{\rm sq}}(U(1))$, which is explicitly
given by
\be
\exp(\H)\stackrel{{\rm sq}}{\simeq} \bigoplus_{n=0}^{\infty} \H_n^{U(\H)}\ot
\ovl{\H}_n^{U(1)}.
\ll{dec1} \ee
Here $\H_n^{U(\H)}$ coincides with  $\H^{\ot n}$, now regarded as the carrier
space of the
\rep\ $\pi_n(U(\H))$, which is, in fact, irreducible for all $n$
\ci{Kir1,Ols1} (also cf.\
subsection 3.3 below). Also, ${\H}_n^{U(1)}$ is just $\C$, but regarded as the
carrier space of
$\pi_n(U(1))$, defined by $\pi_n(z)=z^n$; $\ovl{\H}$ stands for the carrier
space of the conjugate
\rep.

The general context for decompositions of the type (\ref{dec1}) is the theory
of Howe dual pairs
\ci{How1,How3}.    In the present instance, this applies to $\H=\C^k$,  with
$U(k)$ and $U(1)$ being the dual pair in $Sp(2k,\R)$.  (Cf. \ci{Ols4} for the
theory of these pairs
in the infinite-dimensional setting.)

 The construction of the induced space $\F^1$ is effortless in this case.
 The fact that \MW\ reduction
took place at $J=1$ means that the orbit of $U(1)$ in question is the point
$1\in {\bf u(1)}^*$. This
orbit is quantized by the defining \rep\ $\pi_1$ of $U(1)$ on
$\H_1=\C$.   By construction, $\F^1$  is the subspace of $\F\ot
\H_1=\F$ which is invariant under the \rep\ $\pi_R^{-1}\ot \pi_1$. Hence
(\ref{dec1}) implies that $\F^1=\H$.
 The induced \rep\
$\pi^1(U(\H))$ on $\F^1$ is simply the restriction of $\pi_{L,{\rm sq}}(U(\H))$
to this space, so
that $\pi^1 \simeq\pi_{1} $. In other words, we have recovered the defining
\rep.

So far, so good, but unfortunately there  is a  subtlety if one  derives
$\pi_R$ and $\pi_L$
from
geometric quantization. Using the `uncorrected' formalism (as described, e.g.,
in Ch.\ 9 of
\ci{Woo}), exploiting the existence of an invariant positive totally complex
polarization, viz.\
the anti-holomorphic one, one finds that $\F$ is realized as the space of
holomorphic functions on
$\H$. The quantization  $\pq$  of the moment maps $J_R$  for   $U(1)$ and $J_L$
for  $U(\H)$  (with
respect to their respective actions on $\C^k$)  then reproduce the second
quantizations $\pi_{R,{\rm
sq}}$ and $\pi_{L,{\rm sq}}$, respectively.

 If, however, one is more sophisticated and incorporates the  half-form
correction to geometric
quantization \ci[Ch.\ 10]{Woo}, one obtains extra contributions:  for
$\H=\C^k$,  $\pq(J_R)$ is
replaced by  $\pq(J_R)+k/2$, whereas $\pq(J_L)$ acquires an additional constant
$\half$ (times the
unit matrix). These Lie algebra \rep s exponentiate to unitary \rep s of double
covers $\til{U}(k)$
and $\til{U}(1)$, which we denote by $\pi_{L,{\rm
hf}}$ and
$\pi_{R,{\rm hf}}$, respectively. Under
$\pi_{L,{\rm hf}}(\til{U}(k))\ot \pi^{-1}_{R,{\rm hf}}(\til{U}(1))$ we then
find the central
decomposition
\be
\exp(\H)\stackrel{{\rm hf}}{\simeq} \bigoplus_{n=0}^{\infty}
\H_{(n+\half,\half,\ldots,\half)}^{\til{U}(k)}\ot \ovl{\H}_{n+\half
k}^{\til{U}(1)}. \ll{dec2}
\ee
Here $\H_{(n+\half,\half,\ldots,\half)}$ carries the \rep\ of $\til{U}(k)$ with
highest weight $
(n+\half,\half,\ldots,\half)$; this is the tensor product of $\H_n$ and the
square-root of the
determinant \rep. One observes that the inclusion of half-forms is awkward for
Rieffel induction -
we defer a discussion of this point to subsection \ref{discussion}

We finally  turn to the question (discussed in subsection 2.3 in the context of
geometric
quantization)  which observables are quantizable with the Rieffel induction
method (in case it
works, i.e., using $\pi_{L,{\rm sq}}$ and $\pi_{R,{\rm sq}}$!). For
simplicity, in order to have bounded observables we restrict the algebra of
classical observables
$\cin(\H)$ to $C_b^{\infty}(\H)$, and
take the quantization  of the latter to be
the self-adjoint part of the (von Neumann) algebra of all bounded operators
$\B(\F)$ on $\F$. The
subalgebra $\A$ of operators whose Rieffel-induced \rep\ $\pi^1$ on $\F^1$ can
be defined is the
commutant of $\pi_R(U(1))$ - this may be thought of as the quantization of the
Poisson subalgebra
$\sf A$ of $C_b^{\infty}(\H)$ of functions invariant under $U(1)$, i.e.,
satisfying $f(\ps z)=\ps(z)$
for all $z\in U(1)$. (To explain a more intrinsic definition of $\A$ we assume
that the reader is
familiar with the the theory of Hilbert $C^*$-modules and rigging maps, which
play a role in the
general theory of Rieffel induction \ci{Rie74,WO}. $\F$ is a Hilbert
$C^*$-module for the group
algebra $\B=C^*(U(1))$. The rigging map $\la \cdot,\cdot\ra_{\B}$ is given by
\ci{NPL93} $\la
\Psi,\Phi \ra_{\B}(h)= (\pi_R^{-1}(h)\Phi,\Psi)$, which defines a continuous
function on $U(1)$. The
algebra $\A$ is then the algebra of all adjointable maps on $\F$, in that each
$A\in\A$ satisfies $\la
A\Psi,\Phi \ra_{\B}=\la\Psi,A^*\Phi\ra_{\B}$.)

In any case, the von Neumann algebra generated by $\pi_R(U(1))$ consists of the
operators that are
diagonal w.r.t.\ the decomposition $\F=\sum_n \H^{\ot n}$. Hence by a
well-known theorem of von
Neumann stating that the commutant of the algebra of diagonalizable operators
is the algebra of decomposable
operators, $\A$ consists of those bounded operators on $\F$ which map each
$\H^{\ot n}$ into itself.
The induced \rep\ $\pi^1(\A)$ is simply the restriction of $\A$ to $\H$, so
that
$\pi^1(\A)=\B(\H)$. Thus we have not only produced the Hilbert space $\H$ as
the quantization of
$\PH$, but in addition the correct algebra of observables has emerged.

\subsection{The coadjoint orbits of $U_0(\H)$ as reduced spaces}
We recall that $G=U_0(\H)$  is the Banach Lie group of all unitary operators
$U$ on $\H$ for which
$U-{\Bbb I}$ is compact, equipped with the uniform operator (i.e., norm)
topology. Its Lie algebra
$\g={\bf u}_0(\H)=i{\frak K}(\H)_{\rm sa}$ consists of all skew-adjoint compact
operators on $\H$ with
the norm topology. The dual $\g^*={\bf u}_0(\H)^*$ is the space of all
self-adjoint trace-class
operators on $\H$, with topology induced by the trace norm $\n \rh\n_1={\rm
Tr}\, |\rh|$ (this
coincides with the weak$\mbox{}^*$ topology). The pairing is given by
$\la\rh,X\ra=i\, {\rm Tr}\, \rh
X$.

The coadjoint action of $U_0(\H)$ on $\lu$ is given by $\pco(U)\rh=U\rh U^*$.
We are interested in
those coadjoint orbits which are `quantizable' in the sense of geometric
quantization, since
their quantization should produce all irreducible \rep s of $U_0(\H)$
\ci{Kir1,Kir2}.
Each such orbit is labeled by a pair $({\sf m}, {\sf n})$, where ${\sf m}$ is
an ordered $M$-tuple of
positive integers satisfying $m_1\geq m_2\geq\ldots m_M>0$, and $\sf n$ is a
similar $N$-tuple ($M,N<\infty$). The coadjoint orbit $\Omn$ consists of
all elements of $\lu$ with eigenvalues
$m_1,m_2,\ldots,m_M,0^{\infty},-n_N,\ldots,-n_1$. The
degeneracy of each numerical eigenvalue $m_i$ (or $-n_j$) is simply the number
of times it occurs in
this list. The explicit quantization of the orbits $\Omn$ is not discussed in
\ci{Kir1,Kir2};  the
case where either $\sf m$ or $\sf n$ is empty is done in \ci{Boy1} using
geometric quantization.

For finite-dimensional $\H$, it was shown by Montgomery \ci{Mon} that $\O_{{\sf
m},0}$ can be written
as a \MW\ reduced space with respect to the natural right-action   of $U(M)$
on  $\H\ot\C^M$.
This is a special instance of the theory of   dual pairs.
With $\H=\C^k$, the groups $U(\H)$ and $U(M)$ form a Howe dual pair inside the
symplectic group
$Sp(2kM,\R)$ \ci{How1,Ste,How3}, and the moment maps $J_R$ and $J_L$ introduced
below build a
Weinstein dual pair,  cf.\ \ci{KKS,Wei83}.	General theorems on the connection
between coadjoint
orbits of one group and \MW\ reduced spaces w.r.t.\ the other group in a dual
pair are given in
\ci{LMS}. We will now generalize the special case mentioned above to
infinite-dimensional $\H$, and
general orbits $\Omn$.

We take $S=\H\ot\C^{M+N}$, which we regard as a Hilbert manifold in the obvious
way.
We    choose the
canonical basis $\{e_i\}_{i=1,\ldots,M+N}$ in $\C^{M+N}$. The symplectic form
$\om$ on $S$  is taken
as (we put $\pl =1$) \be
\om(\ps,\phv)=-2\, {\rm Im}\, \left(
\sum_{i=1}^M(\ps_i,\phv_i)-\sum_{i=M+1}^{M+N}(\ps_i,\phv_i)\right) , \ll{ommn}
\ee
where we have expanded $\ps=\sum_i\ps_i\ot e_i$ and similarly for $\phv$.
It is convenient to introduce an indefinite sesquilinear form on $\C^{M+N}$ by
putting
$(e_i,e_j)=\pm \dl_{ij}$, with a plus sign for $i=1,\ldots,M$ and a minus sign
for $i=M+1,\ldots,
M+N$. Together with the inner product on $\H$ this induces an indefinite form
$(\cdot,\cdot)_S$
on $S$ in the
obvious (tensor product) way. The r.h.s. of (\ref{ommn}) then simply reads
$-2\, {\rm Im}\,
(\ps,\phv)_S$. A simple trick shows that $S$ is strongly symplectic: we can
regard $S$ as a Hilbert
space $\H\ot\C^M\oplus \ovl{\H\ot\C^N}$,  with   inner product
$(\ps,\phv)_{\footnotesize\rm trick}=\sum_{i=1}^M(\ps_i,\phv_i)+
\sum_{i=M+1}^{M+N}(\phv_i,\ps_i)$.
Then  $\om(\ps,\phv)=-2\, {\rm Im}\,(\ps,\phv)_{\footnotesize\rm trick}$, and
the claim follows from
the well-known fact that Hilbert spaces are strongly symplectic \ci{Mar74,AM}.

 The  Lie group $H=U(M,N)$  (which is $U(M)$ or $U(N)$ for $\sf n$ or $\sf m$
empty) acts on
$S$ from the right in the obvious way, i.e.,  by $U\raw\I\ot U^T$. This action
is symplectic, with
anti-equivariant moment map $J_R:S\raw (\h^*)^-$.  If we identify  $X\in \h$
with a generator in the defining \rep\ of $H$ on $\C^{M+N}$, we obtain (cf.\
\ci[p.\ 501]{KKS}
\be
\la J_R(\ps),X\ra = i(\I\ot X^T\ps,\ps)_S.  \ll{jr}
\ee
On a suitable Cartan
subalgebra $\frak t$ of $\h$, which we identify as the set of  imaginary
diagonal operators on
$\C^{M+N}$,  with basis
$H_j=-iE_{jj}$, this simply reads $\la J_R(\ps),H_j\ra = \pm (\ps_j,\ps_j)$
 with a plus sign for $j=1,\ldots,M$ and a minus sign for $j=M+1,\ldots, M+N$.

We now identify $\wmn$ with an element of $\h^*$ by the pairing $\la
\wmn,X\ra=i{\rm
Tr}\, D_{\wmn}X$, where $ D_{\wmn}$ is the diagonal matrix in $M_{M+N}(\C)$
with entries
\\ $m_1,\ldots, m_M, -n_N,\ldots,-n_1$. This means that $\wmn$ defines a
dominant integral weight on
$\frak t$, and vanishes on its complement.
The subset $J_R^{-1}(\wmn)$ of $S$ is easily seen to consist of those
$\ps=\sum_i\ps_i\ot e_i$
for which $$(\ps_1,\ps_1)=m_1,\ldots,
(\ps_M,\ps_M)=m_M,(\ps_{M+1},\ps_{M+1})=n_N,\ldots,
(\ps_{M+N},\ps_{M+N})=n_1,$$ and $(\ps_i,\ps_j)\sim  \dl_{ij}$. The
normalizations come from
$J_R$ evaluated on
$\frak t$, and the orhtogonality derives from the constraint that $J_R$ vanish
on its complement.
{\em Note that the integrality of the $m_i$ and $n_j$ plays no role in this
subsection.}
\begin{lemma}
$J_R^{-1}(\wmn)$ is a submanifold of $S$. \ll{subm}
\end{lemma}
{\em Proof.}  According to the theorem on p.\ 550 of \ci{AMP}, we   need to
show that $J_R:
J_R^{-1}(\wmn)\raw \h^*$ is a submersion, which is the case if at any point
$\ps\in J_R^{-1}(\wmn)\subset S$ the derivative $(J_R)_*\equiv
J_R^{(1)}:T_{\ps}S\raw
T_{J_R(\ps)}\h^*\simeq \h^*$ is surjective and has a complementable kernel.
The former is equivalent to the statement that $\ps$ is a regular value of the
moment map \ci{AM}.
The derivative at $\ps\in S$ follows from (\ref{jr}) as
 \be
\la (J_R^{(1)})_{\ps}(\xi),X\ra= 2 {\rm Re}\, (\I\ot iX^T\xi,\ps)_S. \ll{derjr}
\ee
 This formula shows that $J_R^{(1)}$ is continuous, so that its kernel is
closed. The
complementability of this kernel is then immediate, since $S$ is a Hilbert
manifold.
The surjectivity of $J_R^{(1)}$ follows from (\ref{derjr}) by inspection, but
it is more instructive
to derive it from Prop.\ 2.11 (due to Smale) in \ci{Mar}. This states that
$\ps$ is a regular value
of the moment map iff the stability group $H_{\ps}\subseteq H$ of $\ps$ is
discrete.
Now, as pointed out earlier,  $\ps=\sum_i \ps_i\ot e_i \in J_R^{-1}(\wmn)$
implies that all
$\ps_i$ are nonzero are orthogonal, so that $H_{\ps}$ is just the identity.
\hfill $\blacksquare$

The action of $H$ on $S$ is not proper unless $\sf m$ or $\sf n$ is empty (in
which case $H$ is
compact). However: \begin{lemma}
The action of $H$ on $J_R^{-1}(\wmn)$ is proper.
\end{lemma}
{\em Proof.} Let $\ps^{(n)}\raw \ps$ in $S$; equivalently, $\ps_i^{(n)}\raw
\ps_i$ in $\H$ for all
$i$. If $\{U^{(n)}\}$ is a sequence in $H$ and $U^{(n)}\ps^{(n)}$ converges,
the fact that for each
$n$ all $\ps_i^{(n)}$ are nonzero and orthogonal implies that
$\{U_{ij}^{(n)}e_j\}$ must converge in
$\C^{M+N}$ for each $i$. Since convergence in the topology on $U(M,N)$ is given
by convergence of
all  matrix elements in the defining representation, this implies that
$\{U^{(n)}\}$ must converge
in $H$. \hfill $\blacksquare$

By the standard theory of \MW\ reduction \ci{Mar74,AM}, these lemmas imply that
the reduced space
\be
S^{\wmn}=J_R^{-1}(\wmn)/H_{\wmn} \ll{sred}
\ee
 (where $H_{\wmn}$ is the stability group of $\wmn\in\h^*$
under the coadjoint action) is a smooth symplectic manifold. We will proceed to
show that it is
symplectomorphic to  the coadjoint orbit $\Omn\in\g^*$, where $G=U_0(\H)$, as
explained above.
The required diffeomorphism is   given by a quotient of the moment map
$J_L:S\raw \g^*$  defined from
the natural left-action of $G$ on $S$, which action is evidently symplectic.
Identifying  $\g$ with
the space of compact skew-adjoint operators $Y$ on $\H$, one easily finds that
this moment map is
given by \be
-i\la J_L(\ps),Y\ra = (Y\ot {\Bbb I}   \ps,\ps)_S
=\sum_{i=1}^M(Y\ps_i,\ps_i)-\sum_{i=M+1}^{M+N}(Y\ps_i,\ps_i).  \ll{jl}
\ee
Since the left-$G$ action and the right-$H$ action commute, $J_L$ is invariant
under $H$ (i.e.,
$J_L(\ps U)=J_L(\ps)$ for all $U\in H$ and $\ps\in \H$), so that $J_L$
(restricted to
$J_R^{-1}(\wmn)$) quotients to a well-defined map $\jlt:S^{\wmn}\raw\Omn$. Once
we have shown that
$\jlt$ is a diffeomorphism, it will follow that it is symplectic, because of
the
 definition
of the  symplectic structure on  $S^{\wmn}$  and the fact that
 $J_L$ is equivariant.

 Generalizing a standard result in the root and weight theory for compact Lie
groups, see e.g.\ \ci{BE},  we first note that
 the  the stability group of $\wmn\in \h^*$ under the coadjoint
action is $H_{\wmn}=\prod_l U(l)$, where $\sum l=M+N$, and the product is over
the multiplicities
within either $\sf m$ or $\sf n$ in
$\wmn$; this is a subgroup of $U(M,N)$ in the obvious block-diagonal form.
(For example, if $\wmn =((2,1,1),(2,2,2))$ the stability group is $U(1)\times
U(2)\times U(3)$.)
It then follows from (\ref{jl}) that $\jlt$ is a bijection onto $\Omn$.

\begin{prop}
$\jlt$ is smooth. \ll{s1}
\end{prop}
{\em Proof.} The manifold structure of $\Omn$ is defined by its embedding in
$\g^*$, which is a
Banach space in the trace-norm topology (cf.\ the beginning of this section).
 The smoothness of $\jlt$ then
follows from that of $J_L:J_R^{-1}(\wmn)\raw\g^*$, since the Lie group $H$ acts
smoothly, freely, and
properly on $J_R^{-1}(\wmn)$.

{\em 1. Continuity of $J_L$.}  We prove continuity on all of $S$. As a map
between separable metric
spaces ($S$ is separable because $\H$ is by assumption, and $\g^*$ is separable
because the
finite-rank operators are dense in it), $J_L$ is continuous if $\ps^{(n)}\raw
\ps$ in $S$ implies
$J_L(\ps^{(n)})\raw J_L(\ps)$ in $\g^*$. The topology on $\g^*$ coincides with
the
weak$\mbox{}^*$-topology, so the desired continuity follows from (\ref{jl}),
the boundedness of $Y$,
and Cauchy-Schwartz.

{\em 2.  Existence and continuity of $J_L^{(1)}$.} The derivative of $J_L$ at
$\ps$ is given by
 \be
\la (J_L^{(1)})_{\ps}(\xi),Y\ra= 2 {\rm Re}\, \left(
\sum_{i=1}^M(iY\xi_i,\ps_i)-\sum_{i=M+1}^{M+N}(iY\xi_i,\ps_i)\right).
\ll{derjl}
\ee
By the same reasoning as in the previous item, $(J_L^{(1)})_{\ps}$ lies in
${\cal L}(S,\g^*)$
and is continuous.

The second derivative $J_L^{(2)}:S\times S\raw \g^*$ can be read off from
(\ref{derjl}); its
existence and continuity are  established as before. Higher derivatives vanish.
\enp

\begin{prop}
$\jlt^{-1}$ is smooth. \ll{s2}
\end{prop}
{\em Proof.} We pick an arbitrary point $\rh_0\in\Omn$, with stability group
$G_0$. Let
$\H=\oplus_l\H_l$ be the decomposition of $\H$ under which $\rh_0$ is diagonal
(the dimension of
each $\H_0$ is the degeneracy of the corresponding eigenvalue; this dimension
is finite unless the
eigenvalue is 0). Then $G_0=\oplus_l U_0(\H_l)$, in self-evident notation. The
Lie algebra $\g_0$
of $G_0$ is given by those operators in $\g=
i{\frak K}(\H)_{\rm sa}$ which commute with $\rh_0$. The manifold $\Omn$ is
modelled on $\g/\g_0$. This has the quotient topology inherited from $\g$,
i.e., the trace-norm
topology determined by $\n A\n_1={\rm Tr}\, |A|$.

We define a neighbourhood $V_0\subset \Omn$ of $\rh_0$ as follows. Since $G$ is
a Banach-Lie group, by
\ci{LT} there exists a neighbourhoud $V$ of $0\in\g$ such that $\exp$ is a
diffeomorphism on $V$ into
$\g$.  We put $V_0=\{\pco(\exp(A))\rh_0| A\in V\}$ (recall that the coadjoint
action is given
by $\pco(U)\rh=U\rh U^*$). To define a chart on $V_0$, we first show that $\g$
(equipped with the
  trace-norm topology) admits a splitting $\g=\g_0\oplus \m_0$.
Here $\m_0$ consists of those operators $A$ in $\g$ whose matrix elements
$(A\ps,\phv)$ vanish
if both $\ps$ and $\phv$ lie in the same space $\H_l$, for all $l$. It is clear
that
 $\g=\g_0\oplus \m_0$ as a set, and it quickly folows  that each summand is
closed:
 since $\n A\n\leq \n A\n_1$, the uniform topology
is weaker than the trace-norm one, so that closedness in the former implies the
corresponding
property in the latter topology.
As to the uniform closedness of  $g_0$, one has $\n [A,\rh_0]\n\,\leq 2 \n
A\n\; \n \rh_0\n$, so
that $\g_0\ni A_n\raw A$   implies that $A\in \g_0$. On $\m_0$ an even more
elementary inequality
does the job. Thus $\g/\g_0\simeq \m_0$,  and we may use $\m_0$ as a modelling
space for $\Omn$.

We define a chart on $V_0$ by $\phv_0:V_0\raw \m_0$, given by
$\phv_0(\pco(\exp(A))\rh_0)=A_0$, where
$A_0$ is the component of $A\in\g$ in $\m_0$.
We would like to  model $S^{\wmn}$ on $\m_0$ as well, but this is not directly
possible
because it has the wrong topology. Hence the following detour.
Take a $\ps_0\in J_R^{-1}(\wmn)\subset S$ for which $J_L(\ps_0)=\rh_0$.
Using the fact that $J_L$ is a  bijection, we  model
$S^{\wmn}=J_R^{-1}(\wmn)/H_{\wmn}$ on the closed
linear   subspace  of $S$ given by $M_0=\{A\ot\I \ps_0|A\in \m_0\}$, equipped
with
  the relative topology of $S$.
Put
$W_0=\{\exp(A)\ot\I \ps_0|A\in m_0\}\subset S$. If
$pr:J_R^{-1}\raw J_R^{-1}(\wmn)/H_{\wmn}$ is the
 canonical projection, we have a chart on the neighbourhood $pr(W_0)$
of $pr(\ps_0)$ defined by  $\ph_0:pr(W_0)\raw M_0$ given by
$\ph_0(pr(\exp(A)\ps_0))=A\ps_0$.
This procedure respects  the  manifold structure of  $S^{\wmn}$, which by
definition is quotiented from $J_R^{-1}(\wmn)\subset S$.

We now define $\mbox{}_0\jlt^{-1}=\ph_0\circ\jlt^{-1}\circ\phv_0^{-1}$; this is
a map from $\phv_0
(V_0)\subset \m_0$ to $\ph_0\circ pr(W_0)\subset M_0$. Clearly,
$\mbox{}_0\jlt^{-1}(A)=A\ps_0$.
This immediately implies that $\mbox{}_0\jlt^{-1}$, and therefore $\jlt^{-1}$,
is smooth.
\enp

It would have been possible to prove Proposition \ref{s1} using the method of
proof of Proposition
\ref{s2}, but that would necessitate an argument (more complicated than our
direct proof of
Proposition \ref{s1})
to the effect
that the trace-norm topology restricted to $\m_0$ is equivalent to the strong
operator topology
\ci{Bon}.  In contrast, in Proposition \ref{s2} we merely needed the continuity
of the identity map
on $\m_0$, with the trace-norm topology as the initial one, and the strong
operator topology as the
final one. This is trivial, for the trace-norm topology is finer than the
uniform topology, which in
turn is finer that the strong operator topology. To sum up, we have proved
\begin{theorem}
For any separable Hilbert space $\H$, the coadjoint orbit $\Omn$ of the group
$U_0(\H)$ (which
consists of all trace-class operators on $\H$ with $M$ specific positive and
$N$ specific negative
 eigenvalues) is
symplectomorphic to the \MW\ quotient $S^{\wmn}=J_R^{-1}(\wmn)/H_{\wmn}$
with respect to $S=\H\ot\C^{M+N}$ and the natural right-action of $H=U(M,N)$.
\ll{omw}
\end{theorem}
\subsection{Representations induced from $U(M)$}
The \rep s of $U_0(\H)$ were fully classified in \ci{Kir1,Ols1,Ols3} (also cf.\
\ci{Kir2,Ols4,Boy2}).
A remarkable fact  is that $U_0(\H)$ is a type I group, so that all its
factorial \rep s are of the
form $\pi\ot\I$ on $\H_{\pi}\ot\H_{\rm mult}$, where $(\pi,\H_{\pi})$ is
irreducible.
Each \irep\ corresponds to an integral weight $\wmn$ of the type specified
above, where $M$ and $N$
are arbitrary (but finite). The  carrier space $\Hmn$ is of the form
$\Hmn=\Hm\ot \Hn$, and carries
the \irep\ $\pmn=\pim\ot\pin$. Here $\Hm$ is the subspace of $\ot^M\H$ obtained
by symmetrization
according to the Young  diagram whose $k$-th row has length $m_k$, and $\Hn$ is
the conjugate space of
$\H^{{\sf n}}$. The \rep\ $\pim$ is the one given by the restriction of the
$M$-fold tensor product
of the defining \rep\ of $U_0(\H)$ to $\Hm$, etc.

 This is almost identical to the theory for
finite-dimensional $\H=\C^k$ \ci{Wey,Zel} (which has the obvious restriction
that
$M,N\leq k$); the only difference
is that in the infinite-dimensional case $\Hm\ot\Hn$ is already irreducible.
For $k<\infty$, on the
other hand, one needs to take  the so-called Young product \ci{Zel} of $\Hm$
and $\Hn$ rather than the
tensor product (this is the irreducible
subspace generated by the tensor product of the highest-weight vectors in each
factor); moreover,
the use of conjugate spaces may be avoided in that case by tensoring with
powers of the determinant
\rep. For example, $\C^k\ot \ovl{\C}^k$ contains the irreducible subspace
$\sum_{i=1}^k e_i\ot
\ovl{e_i}$ which does not lie in the Young product; for $k=\infty$ this
subspace evidently no longer
exists.
 For $M=0$ or $N=0$ there is no difference whatsoever.

We will now show how the \rep s $(\pim,\Hm)$ can be obtained by Rieffel
induction; the
\rep s $(\pin,\Hn)$ may then be constructed similarly. This will quantize the
coadjoint orbits
$\O_{\sf m}\equiv \O_{({\sf m},\emptyset)} $ and $\O^-_{\sf
n}\equiv\O_{(\emptyset,{\sf n})}$,
respectively.
 We note that $\O^-_{\sf n}$ is $\O_{\sf n}$ with the sign of the symplectic
form changed; this
relative minus sign corresponds to the passage from $\H$ to $\ovl{\H}$ upon
quantization.

Our starting point is Theorem \ref{omw}, in which we take $S=\H\ot\C^M$, with
$H=U(M)$ acting on $S$
from the right and $G=U_0(\H)$ acting from the left in the natural way; we call
these actions
$\pi_1^T(H)$ and $\pi_1(G)$, respectively. As
explained in part \ref{ri} of the Introduction, we first have to quantize $S$
and the group actions
defined on it. We do so by taking the bosonic second quantization, or symmetric
Fock space,
$\F=\exp(S)$ over $S$ \ci{RS1,Woo}, cf.\ subsection 3.1.
For later use, we equivalently define  this as the subspace of
$\sum_{n=0}^{\infty} \ot^n S$ on which the natural \rep\ of the symmetric group
$S_n$ on $\ot^n S$
acts   trivially for all $n$.

As in the $M=1$ case (cf.\ subsection 3.1) we first investigate the \rep s of
$U_0(\H)$ and $U(k)$ on
$\F$ obtained by second quantization, or equivalenty, by geometric quantization
without the
half-form modification. This goes as follows.
 The groups $H$ and $G$ act on each subspace $\ot^n S$ by the $n$-fold tensor
product of their respective actions on $S$, and these actions restrict to $\F$.
 Thus the actions
$\pi_1^T(H)$ (which we turn into a \rep\ by taking the inverse) and $\pi_1(G)$
on $S$ are quantized
by the unitary \rep s $\Gamma\ovl{\pi}_1(H)$ ($=\pi_{R,{\rm sq}}^{-1}(H)$ in
the notation of
subsection 3.1, and $\pi_R^{-1}(H)$ in that of subsection \ref{ri}) and $\Gamma
\pi_1(G)$
($=\pi_{L, {\rm sq}}(G)$), respectively (note that
$\pi^T_1(h^{-1})=\ovl{\pi}_1(h)$).
 Here $\Gamma$ is the second quantization functor \ci{RS1}. This setup, and the
associated
central decomposition of $\F$ under these group actions, illustrate Howe's
theory of dual pairs
\ci{How1,How2,How3} in
an infinite-dimensional setting, cf.\ \ci{Ols4}.

The   fact that the coadjoint orbit
$\O_{\sf m}$ of $G$ is (symplectomorphic to) the \MW\ quotient of $S$ with
respect to ${\sf
m}\in\h^*$, cf.\ Theorem \ref{omw},  should now be  reflected, or rather
quantized, by constructing
the unitary \rep\ $\pim(G)$ (which according to Kirillov is  attached to
$\O_{\sf m}$)
by Rieffel induction from the \rep\ $\plm(H)$ attached to the orbit through
$\sf m$ in $H$.
Here $\plm(U(M))$ is simply the unitary \irep\ given by the highest weight $\sf
m$; it is realized
on $\Hlm$, which is the subspace of
 $\ot^M \C^M$ obtained by symmetrization
according to the Young  diagram whose $k$-th row has length $m_k$.

As mentioned in subsection \ref{ri}, to find the carrier space of the induced
\rep\ $\pim(G)$ we
merely have to identify the subspace of $\F\ot\Hlm$ which is invariant under
$\Gamma\ovl{\pi}_1\ot\plm(H)$. This is very easy on the basis of the following
well-known facts
\ci{Wey,Zel,How4}:
\begin{enumerate}
\item
The \rep s of the symmetric group $S_n$ are self-conjugate; for any \irep\
$\pll(S_n)$, the tensor
product $\pll\ot\pll$ contains the identity \rep\ once, and $\pll\ot \plla$
does not contain the
identity unless ${\sf l}={\sf l}'$. (Recall that the \irep s of $S_n$ are
labelled by
an $n$-tuple of integers
${\sf l}=(l_1,\ldots,l_n)$, where $l_1\geq l_2\geq \ldots l_n\geq 0$ and
$\sum_i l_i=n$.)
The collection of all such $n$-tuples $\sf l$ forms the dual $\hat{S}_n$.
\item
Any unitary \irep\ $\pll(U(M))$ is  given by an $M$-tuple
${\sf l}=(l_1,\ldots,l_M)$ of positive nondecreasing integers (possibly zero),
as in the preceding
item, or by the conjugate $\ovl{\pll}$ of such a \rep. Then $\pll\ot\ovl{\pll}$
contains the
identity \rep\ once, but the identity does not occur in any $\pll\ot\plla$, or
in any
$\pll\ot\ovl{\plla}$ unless in the latter case ${\sf l}={\sf l}'$.
\item
The defining \rep\ of $S_n$ on $\ot^n \C^M$ commutes with the $n$-fold tensor
product of
the conjugate of the defining \rep\ of $U(M)$, so that one has the central
decomposition
\be
\ot^n \C^M \simeq \bigoplus_{{\sf l}'\in \hat{S}_n} \Hll^{S^n}\ot \ovl{\H}_{\sf
l}^{U(M)}, \ll{cd1}
\ee
where the prime (relevant only when $M<n$) on the $\oplus$ indicates that the
sum is only over those
$n$-tuples $\sf l$  for which $l_{M+1}=0$. Here $\Hll^{S^n}$ is the carrier
space of $\pll(S^n)$,
and $\ovl{\H}_{\sf l}^{U(M)}$ is  the carrier space of the conjugate of the
\irep\ of $U(M)$ obtained
by making $\sf l$ an $M$-tuple by adding or removing zeros. (A simliar
statement holds without
the conjugation, of course.)
\item
Similarly,
\be
\ot^n \H \simeq \bigoplus_{{\sf l}\in \hat{S}_n} \Hll^{S^n}\ot \Hm, \ll{cd2}
\ee
 under the appropriate \rep s of $S_n$ and $U_0(\H)$, where $\Hm$ was
introduced at the beginning of
this subsection (for $\H=\C^k$ this is equivalent to a classical result in
invariant theory, see e.g.
\ci[4.3.3.9]{How4}).  \end{enumerate}

 Now consider $\ot^n (\H\ot\C^M)\simeq \ot^n \H\,\ot\, \ot^n \C^M$. This
carries the \rep\
$\pi_n^{\H}\ot\pi_n^{\C^M}$ of $S_n$, where $\pi_n^{\cal K}(S_n)$ is the
natural
\rep\ on $\ot^n {\cal K}$. Applying items 4 and 3, and subsequently 1 above, we
find
that the  subspace $\ot^n_s(\H\ot\C^M)\subset \ot^n (\H\ot\C^M)$ which is
invariant under $S_n$ can be
decomposed as
\be
\bigotimes^n_s(\H\ot\C^M)\simeq \bigoplus_{{\sf l}'\in \hat{S}_n} \H^{\sf l}\ot
\ovl{\H}_{\sf
l}^{U(M)}, \ll{decohil}
 \ee
in the sense that the restriction $\ot^n_s (\pi_1(G)\ot\ovl{\pi}_1(H))$
of $\Gamma \pi_1(G)\ot \Gamma\ovl{\pi}_1(H)$ (defined on
$\F=\exp(\H\ot\C^M)$) to $\ot^n_s(\H\ot\C^M)\subset \F$ decomposes as
\be
\bigotimes^n_s (\pi_1(G)\ot\ovl{\pi}_1(H))\simeq \bigoplus_{{\sf l}'\in
\hat{S}_n} \pi^{\sf l}(G)\ot
\ovl{\pll}(H). \ll{decorep}
\ee
 We then apply item 2 to conclude that the only subspace of $\F\ot\Hlm$ which
is
invariant under
$\Gamma\ovl{\pi}_1\ot\plm(H)$ corresponds to $n=\sum_{i=1}^M m_i$ (where $m_i$
are the entries of the
$M$-tuple $\sf m$). Moreover, by (\ref{decohil}) this invariant subspace is
exactly $\Hm$ as a
$U_0(\H)$ module. Hence we have proved
\begin{theorem}
Regard the symmetric Fock space $\F=\exp(\H\ot\C^M)$ as a left-module (\rep\
space) of $U_0(\H)$ and a
right-module of $U(M)$ under the second quantization of their respective
natural actions on
$\H\ot\C^M$. Applying Rieffel induction to this bimodule, inducing from the
\irep\ $\plm(U(M))$ (which corresponds to the highest weight ${\sf
m}=(m_1,\ldots,m_M)$),
yields the induced space $\Hm$  carrying the \irep\ $\pim(U_0(\H))$. \ll{main}
\end{theorem}
This, then, is the exact quantum counterpart  of Theorem \ref{omw}, specialized
to ${\sf
n}=\emptyset$.  As remarked earlier, there exists an obvious analogue of
Theorem \ref{main}
for ${\sf m}=\emptyset$, in which  all Hilbert spaces and \rep s occurring in
the construction
are replaced by their conjugates.

  To prepare for the next subsection we will now give a slight reformulation of
the proof.
We start with finite-dimensional $\H=\C^k$, with $k>M$. Classical invariant
theory \ci{How4} then
provides the decomposition of $\exp(S)$ under $\Gamma \pi_1(U(k))\ot \Gamma
\ovl{\pi}_1(U(M))$ as
\be
\exp(\C^k\ot\C^M) \stackrel{{\rm sq}}{\simeq} \bigoplus_{{\sf l}\in D_M}
\H_{\sf l}^{U(k)}\ot
\ovl{\H}_{\sf l}^{U(M)},
\ll{howe}
\ee
where  the sum is over all Young diagrams (or tuples) $D_M$ with $M$ rows or
less, including
the empty diagram. (Note that it would have been consistent with our previous
notation to write
$(\H^{\sf l})^{U(k)}$ for $\H_{\sf l}^{U(k)}$; both stand for the irreducible
\rep\ of $U(k)$ defined
by the Young diagram $\sf l$.
In what follows, we will reserve the notation $\Hul$ for $\Hll(U_0(\H))$, where
$\H=l^2$.) Eq.\
(\ref{howe}) is an illustration of the theory of Howe dual pairs
\ci{How1,How2,How3}: it exhibits a
multiplicity-free central decomposition of $\F=\exp(S)$ under the commuting
actions of $U(k)$ and
$U(M)$ (which form a dual pair in $Sp(2kM,\R)$, of which   $\F$ carries the
metaplectic \rep).

In order to study the limit $k\raw\infty$ we realize $\exp(\H\ot\C^M)$ (with
$\H=l^2$ now
infinite-dimesional) as an (incomplete) infinite tensor product \ci{vonN} with
respect to the vacuum
vector $\Omega\in\exp(\C^M)$, that is (recalling $\exp(\C^k\ot\C^M)\simeq \ot^k
\exp(\C^M)$),
$\exp(\H\ot\C^M)\simeq \ot_{\Omega}^{\infty}\exp(\C^M)$, where the right-hand
side is the Hilbert
space closure (with respect to the natural inner product on tensor products) of
the linear span of
all vectors of the type $\ps_1\ot\ldots \ps_l\ot\Omega\ot\Omega\ldots$,
$\ps_i\in\exp(\C^M)$, in which
only finitely many entries differ from $\Omega$. (The term `incomplete' refers
to the fact that only
`tails' close to an infinite product of $\Om$'s appear.)
Thus $\exp(\C^k\ot\C^M)\simeq \ot^k \exp(\C^M)$ is naturally embedded in
$\exp(\H\ot\C^M)$ by simply
adding an infinite tail of $\Om$'s, and this provides an embedding
 $\exp(\C^k\ot\C^M)\subset  \exp(\C^{k+1}\ot\C^M)$ as well. Clearly,
$\exp(\H\ot\C^M)$ coincides with the closure of the inductive limit
$\cup_{k=1}^{\infty}
\exp(\C^k\ot\C^M)$ defined by this embedding.

 Choosing the natural basis in $\H=l^2$, we obtain an
embedding $U(k)\subset U(k+1)$, with corresponding actions on $\H$; our group
$U_0(\H)$ (realized
in its defining \rep\ on $\H$) is the norm-closure of the inductive limit group
$\cup_{k=1}^{\infty}
U(k)$. Using the explicit realization of $\H^{\sf l}$ as a Young-symmetrized
tensor product, we
similarly obtain embeddings $\Hll(U(k))\subset \Hll(U(k+1))$. Thus the
inductive limit
 $\cup_{k=1}^{\infty} \Hll(U(k))$ is well-defined. Using (\ref{howe}), we then
have that
$\exp(\H\ot\C^M)$ is the closure of $\cup_{k=1}^{\infty} \oplus_{{\sf l}\in
D_M} \H_{\sf
l}^{U(k)}\ot  \ovl{\H}_{\sf l}^{U(M)}$, which in turn coincides with the
closure of
$\oplus_{{\sf l}\in D_M}\cup_{k=1}^{\infty}\H_{\sf
l}^{U(k)}\ot  \ovl{\H}_{\sf l}^{U(M)}$. We now use the  fact  that
the closure of $\cup_{k=1}^{\infty}\H_{\sf l}^{U(k)}$ is $\Hul$ as a \rep\
space of $U_0(\H)$
(this is obvious given the explicit realization of these spaces, but it is a
deep result that an
analogous fact holds for all \rep s of $U_0(\H)$ \ci{Ols1,Ols3,Ols4}).
This yields the desired decomposition
\be
\exp(\H\ot\C^M) \stackrel{{\rm sq}}{\simeq} \bigoplus_{{\sf l}\in D_M} \H_{\sf
l} \ot
\ovl{\H}_{\sf l}^{U(M)},
\ll{howeinf}
\ee
 under $\Gamma \pi_1(U_0(\H))\ot \Gamma \ovl{\pi}_1(U(M))$. This result was
previously derived in
\ci{Ols4} using a technique of holomorphic extension of \rep s.

Starting from (\ref{howeinf}), Theorem \ref{main} follows immediately from item
2 on the list of
ingredients of our previous proof.

To end this subsection we register how the half-form correction to geometric
quantization modifies
(\ref{howe}), cf.\ subsection 3.1, and in particular (\ref{dec2}). These
corrections are finite only
for $\H=\C^k$, $k<\infty$, so we only discuss that case.
As for $M=1$, one finds that the half-form quantizations of the moment maps
corresponding to the
$U(k)$ and $U(M)$ actions on $\C^k\ot \C^M$ lead to Lie algebra \rep s that can
only be
exponentiated to \rep s $\pi_{L,{\rm hf}}$ and $\pi^{-1}_{R,{\rm hf}}$ of the
covering groups
$\til{U}(k)$ and $\til{U}(M)$ of $U(k)$ and $U(M)$, respectively,
 on which the square-root of the determinant is defined.
 A straightforward exercise
leads to the decomposition
 \be
 \exp(\C^k\ot\C^M) \stackrel{{\rm hf}}{\simeq} \bigoplus_{{\sf l}\in D_M}
\H_{{\sf
l}+\half M}^{\til{U}(k)}\ot  \ovl{\H}_{{\sf l}+\half k}^{\til{U}(M)}
\ll{howehf}
\ee
under $\pi_{L,{\rm hf}}(\til{U}(k))\ot \pi^{-1}_{R,{\rm hf}}(\til{U}(M))$.
Here ${\sf l}+\half M$, regarded as a highest weight, has components
$(l_1+\half M, l_2+\half
M,\ldots)$, and analogously for ${\sf l}+\half k$.  Hence $\H_{{\sf
l}+\half M}$ carries the tensor product of the \rep\ of $\til{U}(k)$
characterized by the Young
diagram $\sf l$, and the determinant \rep\ to the power $M/2$, etc.
 This will be
further discussed in subsection \ref{discussion}.
\subsection{Representations induced from $U(M,N)$}
We  are now going to attempt to `quantize' Theorem \ref{omw} for $N\neq 0$.
The first problem
is finding a unitary \rep\ of $H=U(M,N)$ that corresponds to the dominant
integral  weight $\wmn$ on
$\frak t$ (or the corresponding coadjoint orbit in $\h^*$, cf.\ subsection
3.2); this is the \rep\ we
should induce from.  This problem  was solved in \ci{Ada1}, partly on the basis
of the classification
of all unitary highest-weight modules of $U(M,N)$ \ci{EHW,Jak,Ols2}. In the
compact case, each
dominant integral weight corresponds to an irreducible unitary \rep\ with this
weight as its highest
weight. For $U(M,N)$ on the other hand, there are two new phenomena. Firstly,
there are further
conditions on the dominant integral weight $\wmn$, namely that all entries of
$\sf m$ should be
different, and that all entries of $\sf n$ should be different. Secondly, the
\rep\ corresponding to
$\wmn$, albeit a highest weight \rep, does not in fact have $\wmn$ as its
highest weight. Rather, the
highest weight corresponding to $\wmn$ is `renormalized': with
$m_1>m_2>\ldots>m_M>0$ and
$n_1>n_2>\ldots>n_N>0$, the highest weight (naively expected to be
$(m_1,\ldots,m_M,-n_N,\ldots,-n_1)$)
is in fact  $$(m_1 +\half(N-M)+\half,\ldots,m_i+\half(N-M)+i-\half,\ldots,
m_M+\half (N+M)-\half,$$ $$
-(n_N+\half(M+N)-\half),\ldots,-(n_j+\half(M-N)+j-\half),\ldots,
-(n_1+\half(M-N)+\half))
.$$
Note that this highest weight is still dominant; however, it may no longer be
integral, so that it
defines a projective \rep\ of $U(M,N)$ (single-valued on its double cover
$\til{U}(M,N)$).
These highest weight \rep s belong to the holomorphic discrete series of
$U(M,N)$ \ci{Kna}.

The second problem is the quantization of $S=\H\ot\C^{M+N}$, with the
corresponding actions of
$G=U_0(\H)$ and $H=U(M,N)$.
One regards $U(M,N)$ as a subgroup of $Sp(2(M+N),\R)$, so that the symplectic
action of the former
on $\C^{M+N}$ is the restriction of the action of the latter \ci{Ste,KKS}.
Due to the special way we defined the $U(M,N)$ action in subsection 3.2 as the
inverse of a
right-action, the quantization of this action of $Sp(2(M+N),\R)$ is then
 given by the conjugate of the
metaplectic  \rep\ $\pi_m$ on $L^2(\R^{M+N})\equiv {\cal L}$, cf.\
\ci{KV,SW,Ste}.
This defines a \rep\ of the inverse image $\til{U}(M,N)$ of $U(M,N)$ in the
metaplectic group
$Mp(2(M+N),\R)$ on $\ovl{\cal L}$, which descends to a projective \rep\  of
$U(M,N)$, which we
denote by
$\pi_{{\rm hf}}(\til{U}(M,N))$. As pointed out in \ci{SW} and \ci{BR} (for
$k=1$), this \rep\
is precisely the one obtained from geometric quantization (in a suitable
cohomological variant) if
half-forms are taken into account.
 This yields a first candidate for the quantization of the $U(M,N)$ action on
$\C^{M+N}$.

The second possibility is to take the tensor product of the (restriction of)
the metaplectic \rep\
of  $\til{U}(M,N)$ with the square-root of the determinant, which does define a
unitary \rep\
$\pi_{{\rm sq}}$ of $U(M,N)$ \ci{SW}; see \ci{BR} for a construction of this
\rep\  from
geometric quantization. It is the \rep\ which might be thought of as being
defined by the
physicists' second quantization on $\exp(\C^{M+N})$, as in the $U(M)$ case.
However, since the action
of $U(M,N)$ on $\C^{M+N}$ is not unitary, this second quantization is not, in
fact, defined. In geometric
quantization this lack of unitarity shows up through the non-existence of a
totally complex invariant
polarization on $S$ which is positive. Consequently, one needs to work with an
indefinite such
polarization \ci{BR}, and this leads to complications that will eventually
cause a shift in the \rep s
one would naively expect to occur in the decomposition of the quantization of
$S$.

 For finite-dimensional $\H=\C^k$ we therefore have a suitable quantization of
$S=\C^k\ot\C^{M+N}$,
namely the Hilbert space $\Lk\equiv \ot^k\ovl{\cal L}$ (the Fock space
realization of this space is
not useful, so we drop the notation $\F$). Moreover, we have natural unitary
\rep s
$\ot^k \pi_{{\rm sq/hf}}$ of $\til{U}(M,N)$ on $\Lk$, which are quantizations
of the symplectic
action of $U(M,N)$ on $S$. Following our notation for $U(M)$, we refer to these
\rep s as
$\pi^{-1}_{R,{\rm sq/hf}}$.

 In addition, the quantization of the $U(k)$ action on $S$ may be found (much
more easily) from
geometric quantization with or without half-forms.
The latter case, in which we call the \rep\ $\pi_{L,{\rm sq}}(U(k))$, is
explicitly given in
   \ci{KV}.  Its half-form variant  $\pi_{L,{\rm hf}}(U(k))$
differs from it by the determinant \rep\ raised to the power $(M-N)/2$.

 It follows from the theory of Howe
dual pairs \ci{How1} that $\Lk$ decomposes discretely under these \rep s.
Starting with $\pi_{L,{\rm sq}}(U(k))\ot \pi^{-1}_{R,{\rm sq}}(U(M,N))$,
 the explicit decomposition of $\Lk$ is given in \ci{KV}
as (remember that we have to take the conjugate of the $U(M,N)$ modules, but
not of the $U(k)$
modules used in \ci{KV}, since our $U(k)$ action is the usual one; also, we use
the conventions of
\ci{Ada1} and \ci{How2} for labelling the highest weight, rather than those of
\ci{KV} - this
corresponds to an interchange of $\sf m$ and $\sf n$)
\be
\Lk \stackrel{{\rm sq}}{\simeq} \bigoplus_{\wmn} \H_{\wmn}^{U(k)}\ot
\ovl{\H}_{({\sf m}+ k,{\sf
n})}^{U(M,N)}, \ll{kave}
\ee
where the sum is over all pairs $\wmn$ as defined before, with zeros allowed,
but neither $\sf m$
nor $\sf n$ allowed to be empty. $\H_{\wmn}^{U(k)}$ as a \rep\ space of $U(k)$
was defined in
subsection 3.3, and $\H_{({\sf m}+k,{\sf n})}^{U(M,N)}$ carries the unitary
\rep\ of $U(M,N)$
with highest weight (not subject to
further `renormalization')
$$(m_1+k,\ldots,m_i+k,\ldots,m_M+k,-n_N,\ldots,-n_j,\ldots,-n_1).$$

The decomposition under  $\pi_{L,{\rm hf}}(U(k))\ot \pi^{-1}_{R,{\rm
hf}}(U(M,N))$, on the other
hand, reads \ci{How2}
\be
\Lk \stackrel{{\rm hf}}{\simeq} \bigoplus_{\wmn}
 \H_{({\sf m}+\half(M-N),{\sf n}-\half(M-N))}^{\til{U}(k)}\ot
\ovl{\H}_{({\sf m}+ \half k,{\sf n} +\half k)}^{\til{U}(M,N)}, \ll{kave2}
\ee
 where the  highest weight $({\sf m}+ \half k,{\sf n} +\half k)$ is explicitly
given by

$$(m_1+k/2,\ldots,m_i+k/2,\ldots,m_M+k/2,-n_N-k/2,\ldots,n_j-k/2,\ldots,-n_1-k/2),$$
whereas $\H_{({\sf m}+\half(M-N),{\sf n}-\half(M-N))}$ is the tensor product of
$\H_{\wmn}$,   and $\C$, carrying
the determinant \rep\ of $U(k)$ to the
power $(M-N)/2$, cf.\ \ci{How2}).

Working with (\ref{kave} for the sake of concreteness,
we now wish to apply Rieffel induction from a suitable \rep\ of $H=U(M,N)$ to
$\Lk$ in order to
extract the copy of $\H_{\wmn}^{U(k)}$ for the value of $\wmn$ selected by the
\rep\ we induce from.
Firstly, we need a dense subspace $L\subset \Lk$ such that the
function $x\raw ( \pi^{-1}_{R,{\rm sq}}(x)\ps,\phv)$ is in $L^1(H)$ for all
$\ps,\phv\in L$, cf.\
subsection \ref{ri}. This is easily found: using the decomposition
(\ref{kave}), we take $L$ to
consist of  vectors having a finite number of components in the
decomposition, each component of which is in the tensor product of
$\H^{U(k)}_{\ldots}$ and the dense
subspace of $K$-finite vectors in the other factor. Since each function of the
type $x\raw
(\pi(x)\ps,\phv)$, where $\pi$ is in the discrete series, and $\ps$ and $\phv$
are $K$-finite
vectors, is in Harish-Chandra's Schwartz space \ci{Kna} (which is a subspace of
$L^1(H)$), this
choice indeed satisfies the demand.  (Based on the explicit realization of
$\Lk$ as a function space
\ci{KV}, a more `geometric' choice of $L$ may also be found.)

As we are going to induce from holomorphic discrete series \rep s of $U(M,N)$,
let us examine
the tensor product $\ovl{\H}_{({\sf m}_1,{\sf n}_1)}^{U(M,N)}\ot
\H_{({\sf m}_2,{\sf n}_2)}^{U(M,N)}$.
Recall that $\wmn$
(which here refers to the actual highest weight, rather than the dominant
integral weight that is
subject to renormalization, as sketched above) defines a unitary irreducible
\rep\ $\pi_{\wmn}$ of the
maximal compact subgroup $K=U(M)\times U(N)$ with highest weight
$(m_1,\ldots,m_M,-n_N,\ldots,-n_1)$.
 By Theorem 2 in \ci{Rep}, the above tensor product is unitarily equivalent as
a
\rep\ space of $U(M,N)$ to  the \rep\ induced (in the usual, Mackey, sense)
from
$\ovl{\pi}_{({\sf m}_1,{\sf n}_1)}\ot\pi_{({\sf m}_2,{\sf n}_2)}(K)$. Using the
reduction-induction
theorem, we can therefore decompose this induced \rep\ as a direct sum over the
\rep s induced
from the components in the decomposition of
 $\ovl{\pi}_{({\sf m}_1,{\sf n}_1)}\ot\pi_{({\sf m}_2,{\sf n}_2)}(K)$.

Let us examine a generic \rep\ $\pi^{\kappa}(H)$ (realized on the Hilbert space
$\H^{\kappa}$ of
functions $\ps:G\raw\H_{\kappa}$ satisfying the equivariance
condition $\ps(xk)=\pi_{\kappa}(k^{-1})\ps(x)$) induced from an irreducible
\rep\ $\pi_{\kappa}(K)$. The Rieffel induction procedure produces the
semi-definite form $(\cdot
,\cdot )_0$ on $L\ot\H_{\ch}$ (where, in this case, $\H_{\ch}=\H_{({\sf m},{\sf
n})}^{U(M,N)}$ for
certain $\wmn$). Using (\ref{kave}) and the previous paragraph, we find that
$L\ot\H_{\ch}$ is a
certain dense subspace of a direct sum with components of the type
$\H_{\wmn}^{U(k)}\ot\H^{\kappa}$,
in which $H$ acts trivially on the first factor. By our construction of $L$,
each element of
$L\ot\H_{\ch}$ only has components in a finite number of these Hilbert spaces,
so that we can
investigate each component separately. (Had the number of components of
elements of $L$  been
infinite, the study of $(\cdot ,\cdot )_0$ would have been more involved, as
this is an unbounded and
non-closable quadratic form, so that $(\sum_i\ps_i,\phv)_0\neq
\sum_i(\ps_i,\phv)_0$ for infinite
sums.)

Factorizing $\int_H dx= \int_N dn\,\int_K dk$ \ci{Kna}, it
follows from the equivariance condition  and the orthogonality relations for
compact groups that
in a given component $\H_{\wmn}^{U(k)}\ot\H^{\kappa}$ the expression
$(\ps,\phv)_0=\int_H dx\, ({\Bbb I}\ot\pi^{\kappa}(x)\ps,\phv)$ vanishes unless
$\pi_{\kappa}$ is the
identity \rep\ $\pi_{\rm id}$ of $K$.
Given a highest weight \rep\ $\pi_{\ch}(H)$ we Rieffel-induce from, there
exists a unique pair
$\wmn$ for which $\H_{\wmn}^{U(k)}\ot\H^{\rm id}$ occurs in the decomposition
of $\Lk\ot\H_{\ch}$ as
a sum over induced \rep s of $H$ in the above sense.

 Let
$L^{\rm id}$ be the projection of $L\ot\H_{\ch}$ onto this
$\H_{\wmn}^{U(k)}\ot\H^{\rm id}$.
We define $\til{V}:L^{\rm id}\raw \H_{\wmn}^{U(k)}$ by linear extension of
$\til{V}\ps_1\ot\ps_2=\ps_1\int_Hdx\, \ps_2(x)$ (where $\ps_1\in
\H_{\wmn}^{U(k)}$ and $\ps_2\in\H^{\rm
id}\subset L^2(G)$). The integral exists by our assumptions on $L$; moreover,
the explicit form of
the inner product in $\H^{\rm id}$ (namely $(f,g)=\int_H dx\, f(x)\ovl{g(x)}$,
as $K$ is compact)
leads to the equality $(\til{V}\ps,\til{V}\phv)=(\ps,\phv)_0$ (where the inner
product on the
left-hand side is the one in $\H_{\wmn}^{U(k)}$). We now extend $\til{V}$ to a
map $V$ from
$L\ot\H_{\ch}$ to $\H_{\wmn}^{U(k)}$ by putting it equal to zero on all spaces
involving a factor
$\H^{\kappa}$, where $\kappa\neq {\rm id}$ (and equal to $V$ on $L^{\rm id}$,
of course). Clearly, by
this and the preceding paragraph, \be
(V\ps,V\phv)=(\ps,\phv)_0. \ll{V}
\ee
We are now in a standard situation in the theory of Riefel induction, in which
we can identify the
null space of $(\cdot ,\cdot )_0$ with the kernel of $V$, and the induced space
$\H^{\ch}$ (which, we
recall, is the completion of the quotient of $L\ot\H_{\ch}$ by this null space
in the inner product
obtained from this form) with the closure of the image of $V$. It is clear from
our definition of
$L$ that the image of $V$ actually coincides with $\H_{\wmn}^{U(k)}$. Also, the
definition of the
induced \rep\ $\pi^{\ch}$ of $G=U(k)$ on $\H^{\ch}$ immediately implies that
$\pi^{\ch}\simeq
\pi_{\wmn}$. Finally, note that  (\ref{V}) shows explicitly that
$(\cdot,\cdot)_0$ is positive
semi-definite, a fact which was already certified by Prop.\ 2 in \ci{NPL93}.

Putting these arguments together, we have proved:
\begin{theorem}
Let $U(k)$ and $U(M,N)$ act on $S=\C^k\ot\C^{M+N}$ (equipped with the
symplectic form
 (\ref{ommn})) from the left and
the right, respectively, in the natural way, and let $\Lk$ be the quantization
of $S$, with
commuting \rep s
of  $U(k)$ and $U(M,N)$ on $\Lk$ (which quantize the above symplectic actions)
as given
(up to conjugation of the \rep\ of $U(M,N)$) by
Kashiwara-Vergne \ci{KV}.

Then Rieffel induction on $\Lk$ from the holomorphic discrete series \rep\ of
$U(M,N)$
with highest weight $({\sf m}+ k,{\sf n})$ (that is, the highest weight with
components
$(m_1+k,\ldots,m_M+k,-n_N,\ldots,-n_1)$) leads to an induced space
$\Hlmn^{U(k)}$, which
as a Rieffel-induced $U(k)$ module carries
the \rep\   $\pi_{\wmn}(U(k))$ (which is the Young product of the \rep\ with
Young diagram $\sf m$ and the conjugate of the \rep\ with Young diagram $\sf
n$).

Moreover, the induced space is empty if one induces from a highest weight \rep\
of $U(M,N)$
 of the form $\wmn$ in which at least one $m_i$ is smaller than $k$, or is not
integral. \ll{dis}
 \end{theorem}
\subsection{Discussion}
{The last part of the theorem is particularly unpleasant for the quantization
theory of constrained
system, for it shows that Theorem \ref{omw} cannot really be `quantized' unless
$\sf m$ or $\sf n$
are empty. For we would naturally induce from the holomorphic discrete series
\rep\ of $U(M,N)$
having the `renormalized' highest weight corresponding to a coadjoint orbit
characterized by $\wmn$,
as explained at the beginning of this subsection. But then for $k$ large enough
the induced space will
be empty, rather than consisting of  $\Hlmn^{U(k)}$, as desired. As we have
seen,
the induction procedure is only successful if we induce from a \rep\ with
highest weight $({\sf
m}+k,{\sf n})$, rather than from the ($k$-independent) renormalized weight we
ought to use by
first principles.
This is bizarre,
given that the original weight $\wmn$  (or the orbit it corresponds to) knows
nothing about $k$ or
$U(k)$. In addition, even without this problem the induced space will often be
empty, because the
`correct'
renormalized highest weight one induces from may simply not occur in the
Kashiwara-Vergne
decomposition (\ref{kave}) because of the half-integral nature of its entries
(which is a pure
`quantum' phenomenon).
 (In a rather different setting, the discrepancy for large $k$ between the
`decomposition' of $S$
into  pairs of matched coadjoint orbits for $U(k)$ and $U(M,N)$, and the
decomposition of $\Lk$ under
these groups, must have been noticed by Adams \ci{Ada2},  who points out that
there is a good
correspondence for $k\leq {\rm min}\,(M,N)$ only.)

It is peculiar to the non-compact ($N\neq 0$) case that this difficulty even
arises if the half-form
correction to quantization is not applied.
For (\ref{kave}) is the non-compact analogue of (\ref{howe}), and in the latter
quantization clearly
does commute with reduction. If we do incorporate half-forms, we obtain
(\ref{kave2}) for $U(M,N)$
and (\ref{howehf}) for $U(M)$. In both cases the Rieffel induction process
generically
(that is, if $M\neq N$) fails to produce
the correct \rep\ of $U(k)$, even if one induces from a \rep\ whose highest
weight is renormalized
(compared to the weight expected from the orbit correspondence) by the term
$k/2$.

 Finally, the passage from $\C^k$ to infinite-dimensional
Hilbert spaces is tortuous whenever half-forms are used (the corrections being
infinite for
$k=\infty$), and in the non-compact case even without these.
 This is partly because of the
$k$-dependence of the highest weights of $U(M,N)$, and partly because $\cal L$
does not contain the
identity \rep\ of $U(M,N)$ (recall that in the compact case we used the carrier
space $\C\Omega$ of
this \rep\ as the fixed `tail' vector to construct the von Neumann infinite
tensor product from).

Clearly, this situation deserves further study. We do not think it is an
artifact of our proposal of
using Rieffel induction in the quantization of constrained systems. In fact,
this technique comprises
the only method known to us which is precise enough to bring the embarrassment
to light.
\ll{discussion} }

 \end{document}